# Interplay between Structural Hierarchy and Exciton Diffusion in Artificial Light Harvesting

*Björn Kriete[1], Julian Lüttig[2], Tenzin Kunsel[1], Pavel Malý[2], Thomas L. C. Jansen[1], Jasper Knoester[1], Tobias Brixner[2,3], and Maxim S. Pshenichnikov[1,*]*

[1]University of Groningen, Zernike Institute for Advanced Materials, Nijenborgh 4, 9747 AG Groningen, The Netherlands

[2]Institut für Physikalische und Theoretische Chemie, Universität Würzburg, Am Hubland, 97074 Würzburg, Germany

[3]Center for Nanosystems Chemistry (CNC), Universität Würzburg, Theodor-Boveri-Weg, 97074 Würzburg, Germany

*Abstract*

Unravelling the nature of energy transport in multi-chromophoric photosynthetic complexes is essential to extract valuable design blueprints for light-harvesting applications. Long-range exciton transport in such systems is facilitated by a combination of delocalized excitation wavefunctions (excitons) and remarkable exciton diffusivities. The unambiguous identification of the exciton transport, however, is intrinsically challenging due to the system's sheer complexity. Here we address this challenge by employing a novel spectroscopic lab-on-a-chip approach: A combination of ultrafast coherent two-dimensional spectroscopy and microfluidics working in tandem with theoretical modelling. This allowed us to unveil exciton transport throughout the entire hierarchical supramolecular structure of a double-walled artificial light-harvesting complex. We show that at low exciton densities, the outer layer acts as an antenna that supplies excitons to the inner tube, while under high excitation fluences it protects the inner tube from overburning. Our findings shed light on the excitonic trajectories across different sub-units of a multi-layered supramolecular structure and underpin the great potential of artificial light-harvesting complexes for directional excitation energy transport.

Many natural photosynthetic complexes utilize light-harvesting antenna systems that enable them to perform photosynthesis under extreme low light conditions only possible due to remarkably efficient energy transfer[1]. The success of natural systems, such as the multi-walled tubular chlorosomes of green sulfur bacteria, relies on the tight packing of thousands of strongly coupled molecules[2]. This arrangement facilitates the formation of collective, highly delocalized excited states (Frenkel excitons) upon light absorption as well as remarkably high exciton diffusivities[3]. Understanding the origin of the delocalized states and tracking energy transport throughout the entire complex hierarchy in multi-chromophoric systems – from the individual molecules all the way up to the complete multi-layered assembly – is vital to unravel nature's highly successful design principles.

In reality, however, natural systems are notoriously challenging to work with as they suffer from sample degradation once extracted from their stabilizing environment and feature inherently heterogeneous structures[4–7], which disguises relations between supramolecular morphology and excitonic properties. In this context, a class of multi-layered, supramolecular nanotubes holds promise as artificial light-harvesting systems owing to their intriguing optical properties and structural homogeneity paired with self-assembly capabilities and robustness[8–10]. Previous studies have demonstrated the potential of these systems as quasi-one-dimensional long-range energy transport wires[11–14], where the dependence of the transport properties on the hierarchical order as well as dimensionality of the respective system is a re-occurring topic of great interest[15–17]. Nevertheless, even in these simpler structures the delicate interplay between individual sub-units of the supramolecular assembly hampers the unambiguous retrieval of exciton transport dynamics.

Recent studies have focused on reducing the complexity of multi-layered, supramolecular nanotubes and thereby essentially uncoupling individual hierarchical units of the assembly by oxidation chemistry[9,10,18–20]. In addition, Eisele *et al.* have demonstrated flash-dilution as an elegant tool to selectively dissolve the outer layer to obtain an unobscured view on the isolated inner layer[9,15]. Nevertheless, the rapid recovery of the initial nanotube structure within a few seconds impedes studies more elaborate than simple absorption – for instance, time-resolved spectroscopy – to probe exciton dynamics. A strategy that is capable to alleviate these limitations relies on microfluidics[21], which in recent years has successfully been implemented to manipulate chemical reactions in real time[22] or to steer self-assembly dynamics[23,24]. In particular, combinations of microfluidics and spectroscopy including steady-state absorption[25], time-resolved spectroscopy[26–29], and coherent two-dimensional (2D) infrared spectroscopy[30] have received considerable attention. In this framework, microfluidics bridges the gap between controlled modifications of the sample on timescales of microseconds to minutes with ultrafast processes on timescales down to femtoseconds.

In parallel with these developments, electronic 2D spectroscopy has evolved to a state-of-the-art tool for investigation of exciton dynamics in multi-chromophoric systems with significant inputs from both theory[31–36] and experiment[37–44]. Recently, a fifth-order 2D spectroscopic technique has been demonstrated to be capable of resolving exciton transport properties by directly probing mutual exciton–exciton interactions (hereafter denoted as EEI)[45].

In this paper, we identify the dynamics of excitons residing on different subunits of a multi-walled artificial light-harvesting complex. Disentangling the otherwise complex response is made possible by successfully interfacing EEI2D spectroscopy with a microfluidic platform, which provides spectroscopic access to the simplified single-walled nanotubes. We show that experimental EEI2D spectra, together with extensive theoretical modelling, provide an unobscured view on exciton trajectories throughout the complex supramolecular assembly and allows to obtain a unified picture of the exciton dynamics.

*Results & Discussion*

The absorption spectrum of double-walled C8S3-based nanotubes (Figure 1a, black solid line) comprises two distinct peaks that have been previously assigned to the outer (589 nm, $\omega_{outer}$ ~17000 cm$^{-1}$) and inner layer (599 nm, $\omega_{inner}$ ~16700 cm$^{-1}$) of the assembly[9,19]. The spectral red-shift of ~80 nm (~2400 cm$^{-1}$) and a tenfold spectral narrowing relative to the monomer absorption is typical for J-aggregation[8]. The magnitude of these effects evidences strong intermolecular couplings, which are essential for the formation of delocalized excited states. A number of weaker transitions at the blue flank of the nanotube spectrum were previously ascribed to the complex molecular packing into helical strands[46] with two molecules per unit cell[9]. It has previously been shown that the two main transitions as well as one of the weaker transitions at ~571 nm (~17500 cm$^{-1}$) are polarized parallel, while all remaining transitions are polarized orthogonal to the nanotube's long axis[19]. The nanotubes preferentially align along the flow in the sample cuvette due to their large aspect ratio (outer diameter ~13 nm, length several μm's). As a result, the laser pulses polarized along the flow selectively excite transitions that are polarized parallel to the long axis of the nanotube, i.e., predominantly the two main transitions.

Controlled destruction of the outer layer (Figure 1b) was achieved in a microfluidic flow-cell (Figure 2a) by mixing nanotube solution with a diluting agent (50:50 mixture by volume of $H_2O$ and methanol). Continuous dissolution is evident from the absence of the outer tube absorption peak, while the peak associated with the inner tube is retained (Figure 1a, gray line), which corroborates the 1-to-1 assignment of these peaks to the inner and outer tube. Simultaneously, a new absorption peak around 520 nm (~19200 cm$^{-1}$) indicates an increase in monomer concentration that formerly constituted the outer layer. We use this peak to estimate the concentration of molecules that remains embedded in the inner tubes upon flash-dilution (Supplementary Note 1).

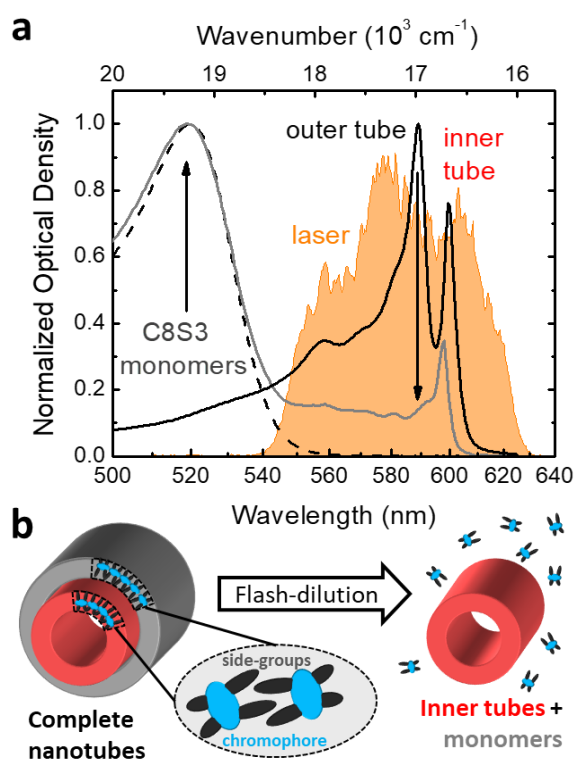

**Figure 1. Absorption spectra before and after flash-dilution.** (**a**) Linear absorption spectra of neat nanotubes (black solid line), isolated inner tubes (gray solid line), and dissolved monomers (black

dashed line) in methanol. The laser excitation spectrum (orange) is shown for comparison. Arrows indicate spectroscopic changes upon flash-dilution. (**b**) Schematic representation of the flash-dilution process that selectively strips the outer tube, while leaving a sufficient share of the inner tubes intact. The decreased amplitude of the peak at ~600 nm indicates partial dissolution of inner tubes. The dissolved monomers contribute to a broad absorption band around ~520 nm, which is not covered by the excitation spectrum and, thus, has no consequences for ultrafast spectroscopy.

A set of representative 2D spectra obtained for complete nanotubes and isolated inner tubes at two waiting times $T$ and the excitation axis expanded to more than twice the fundamental frequency $2\omega$, are shown in Figure 2b. We will refer to the $\omega$ and $2\omega$ regions as absorptive 2D and EEI2D spectra, respectively. It has previously been shown that the $2\omega$ region is dominated by signals that encode exciton–exciton interactions, e.g., exciton–exciton annihilation (EEA)[45,47]. Hence, the structure and dynamics of the EEI2D spectra allow tracing the annihilation of two excitons with their trajectories encoded in the amplitude and spectral position of the respective peak as functions of the waiting time $T$.

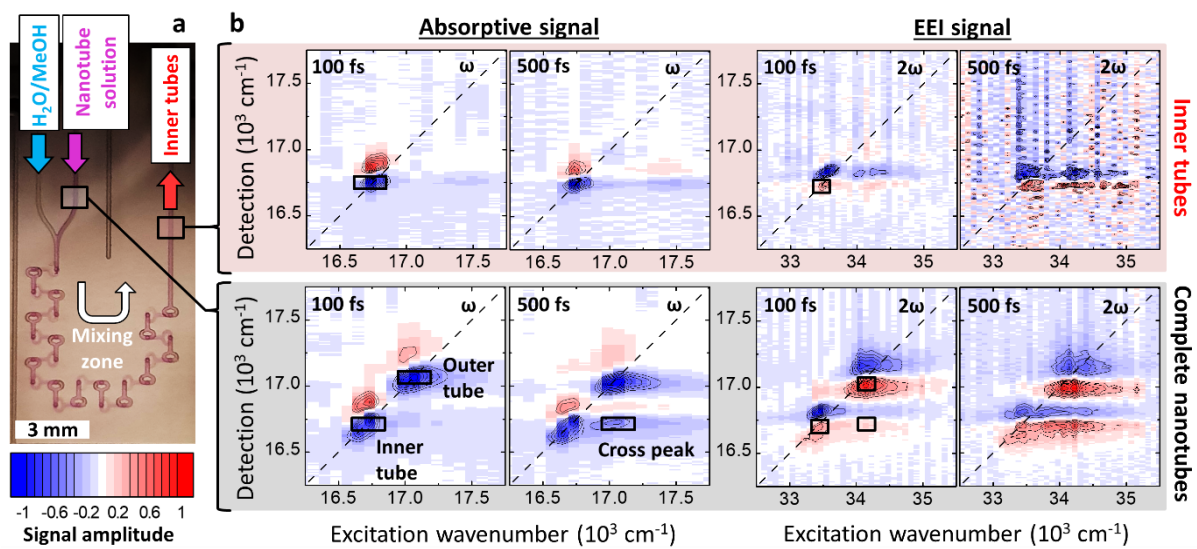

**Figure 2. Absorptive and EEI 2D spectra recorded before and after microfluidic flash-dilution.**
(**a**) Cuvette for microfluidic flash-dilution via mixing of neat nanotube solution and a diluting agent (50:50 mixture by volume of H$_2$O and methanol). Arrows indicate the flow direction of the solvents. (**b**) Representative absorptive 2D and EEI2D spectra at two selected waiting times (100 fs and 500 fs) measured for isolated inner tubes and complete nanotubes in the top and bottom panels, respectively. For better visibility of all peaks, the spectra were normalized to their maximum absolute amplitude. The signal amplitude is depicted on a color scale ranging from -1 to 1, with contour lines drawn at increments of 0.1 except for the lower signal levels. Negative and positive features in the absorptive 2D spectra refer to ground-state bleach/stimulated emission (GSB/SE) and excited-state absorption (ESA) signals, respectively. In the EEI2D spectra the signal signs are opposite, which is caused by the two additionally required interactions with the incident light fields and the associated factor of $i^2 = -1$ within the perturbation expansion[31,45,48]. Diagonal lines (dashed) are drawn at $\omega_{excitation} = \omega_{detection}$ and $\omega_{excitation} = 2\omega_{detection}$ for absorptive 2D and EEI2D spectra, respectively. Black rectangles depict the regions of interest in which the signal was integrated to obtain the transients (Supplementary Table 1). The exciton density corresponds to one exciton per ~20 individual molecules. Spectra for other exciton densities and waiting times are presented in the SI (Supplementary Figure 2 and Figure 3). Note that the direct comparability of the absorptive and EEI signals is ensured, because both signals are recorded under identical conditions, as they are emitted in the same phase-matched direction and captured simultaneously.

For complete nanotubes, the absorptive 2D spectra at early waiting times are characterized by two pairs of negative ground-state bleach/stimulated emission (GSB/SE) and positive excited-state absorption (ESA) diagonal peaks with the low- and high-energy pair associated with the inner tube and outer tube, respectively (Figure 2b, bottom). For later waiting times, a cross peak clearly emerges below the diagonal, for which again GSB/SE and ESA features can be identified; these data are in line with previous publications[15,49]. A cross peak above the diagonal can also be identified; however, it has a low amplitude because of uphill energy transfer and its partial spectral overlap with ESA of the inner tube. The EEI2D spectra essentially mirror the absorptive 2D spectra evidencing intensive exciton–exciton interactions on each individual tube (diagonal peaks) as well as between the tubes (cross peaks).

Upon microfluidic flash-dilution of the outer wall, the 2D spectra simplify to a single pair of GSB/SE and ESA peaks originating from the isolated inner tubes at an excitation frequency of ~16700 cm$^{-1}$ (Figure 2b, top). Expectedly, neither a diagonal peak showing the presence of the outer tube nor a cross peak indicating inter-layer exciton transfer is detected. The absence of the outer tube spectrally isolates weak cross peaks at a detection frequency of ~16700 cm$^{-1}$ and excitation frequencies of ~17500 cm$^{-1}$ and ~35000 cm$^{-1}$ in the absorptive 2D and EEI2D spectra, respectively. These peaks are linked to the blue-shifted transition in the nanotube absorption (Figure 1a) and are not relevant for the further analysis due to their small amplitude (Supplementary Note 4).

In the further analysis, we will focus on the GSB/SE components of the absorptive and EEI signals corresponding to the diagonal outer tube, diagonal inner tube and their low-frequency cross peak, from which we extract the amplitudes as a function of the waiting time for all measured exciton densities by integrating the signal in the rectangles (250 cm$^{-1}$ along the excitation and 100 cm$^{-1}$ along the detection axis; depicted in Figure 2b; Supplementary Table 1). The GSB/SE signals contain information on the creation of excitons residing on different, spatially separated domains followed by EEA due to exciton diffusion.

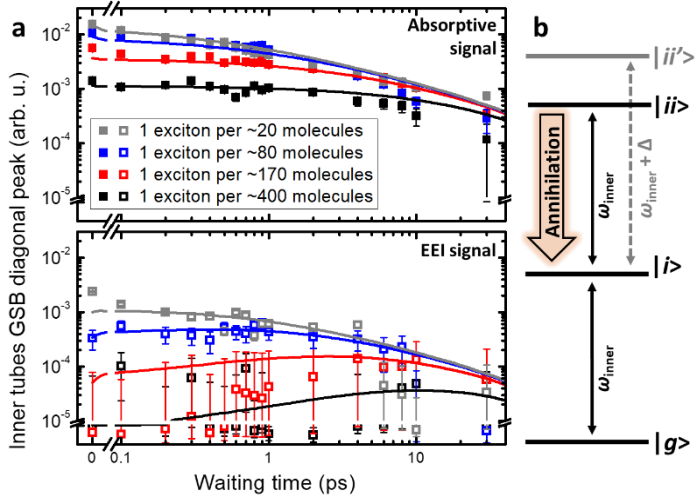

**Figure 3. Absorptive and EEI transients of isolated inner tubes.** (**a**) Log-log plot of the absorptive (upper panel, solid squares) and EEI (lower panel, open squares) GSB/SE transients for isolated inner tubes for different exciton densities. The transients were obtained by integrating the signal in the rectangular regions of interest shown in Figure 2b; the panels are drawn with the same scaling to emphasize their direct comparability, which is one of the constraints in the Monte-Carlo simulations (*vide infra*). The sign of the EEI responses was inverted for the ease of comparison. The error bars refer to the detection noise level in the experiment (Supplementary Note 2). The solid lines depict the results from Monte-Carlo simulations of the exciton dynamics on isolated inner tubes. The amplitude (vertical) scaling between experimental and simulated data is preserved, i.e., for each signal (absorptive and EEI) a single scaling factor was used for *all* simulated transients. (**b**) Energy level diagram of the isolated inner nanotubes with the electronic ground state ($|g\rangle$) and the one- ($|i\rangle$) and bi-exciton ($|ii\rangle$) states (*i* stands for the *i*nner tube). Optical transitions are marked by vertical black arrows with the corresponding frequency $\omega_{\text{inner}}$. The blue-shifted one- to two-exciton transition within the same excited domain ($|ii'\rangle$, dashed gray arrow; Refs. [50,51]) is shown for comparison. Bold arrow: annihilation channel from the bi-excitonic state.

We begin our analysis with the isolated inner nanotubes (Figure 3a). Increasing the exciton density leads to a progressively growing amplitude of the absorptive signal at early waiting times with the onset of saturation at the highest exciton density of 1 exciton per ~20 molecules (Figure 3a, upper panel). Furthermore, the transients decay faster at longer waiting times which is a typical fingerprint for EEA encoded in the EEI signal.

In order to dissect the contributions to the EEI signal, we describe the isolated inner tubes as a three-level system (Figure 3b). The detection frequency selection allows to distinguish between the bi-exciton state of two separate singly-excited domains ($\omega_{inner}$) and the one- to two-exciton transition within the same excited domain ($\omega_{inner} + \Delta$),[50,51] where the latter occurs at a blue-shifted detection frequency as a consequence of Pauli repulsion between excitons[52]. EEA opens a relaxation channel between the $|ii\rangle$ and $|i\rangle$ states[31,34,35,45]. This leads to the re-appearance of the otherwise mutually annulled Feynman diagrams, which in turn results in the emergence of the EEI signal (Supplementary Note 5).

At low exciton densities the EEI signal is barely detectable at the noise background (Figure 3a, black squares), while higher exciton densities lead to the rapid emergence of the EEI signal. For sparse exciton populations a delayed formation of the maximum annihilation signal is glimpsed at a waiting time of ~8 ps (Figure 3a, red squares), because excitons must diffuse towards each other prior to annihilation. This maximum is gradually shifting towards earlier waiting times for higher exciton densities, as a shorter and shorter period is required before individual excitons meet and annihilate. For the highest exciton density, the maximum EEI signal occurs at essentially zero waiting time, as excitons annihilate with virtually no time to diffuse. These features qualitatively agree with predictions of analytical models for diffusion-assisted bi-excitonic annihilation in one and two dimensions[45,53,54]. However, the quantitative description is prevented by the fact that the isolated inner tubes fall in neither category, as the underlying molecular structure shows characteristics of both: helical molecular strands (1D) mapped onto the surface of a cylinder (2D).

We analyze the experimental data using Monte-Carlo (MC) simulations (Methods and Supplementary Note 6), where we describe the exciton dynamics in a combined framework of diffusive exciton hopping and exciton–exciton interactions[45,55–57]. For comparison with experiment, we obtain the amplitude of the absorptive signal by counting the total number of excitons at time $T$ in the MC simulations, whereas for the EEI signal only excitons that have participated in at least one annihilation event are calculated (Supplementary Note 6.2). The latter occurs if two excitons approach each other closer than the annihilation radius, which we define as the cut-off distance for exciton–exciton interactions (Supplementary Note 6.3). We find excellent agreement of the experimental data (Figure 3a, squares) and the simulated curves (Figure 3a, solid lines) by global adjustment of only two parameters: the exciton diffusion of $D_{2D} \sim 5.5$ nm$^2$ ps$^{-1}$ (equivalent to 10 molecules ps$^{-1}$ given the molecular grid in the MC simulations) and the exciton annihilation radius of 3 molecules An overview of all parameters is given in Supplementary Note 6.4. The 2D diffusion constant was obtained via the mean square exciton displacement ($<x^2> = 4D_{2D}\tau$; Supplementary Note 6.5) in the annihilation-free case. Our simulations also revealed that pure two-excitonic annihilation, where each exciton can only participate in a single annihilation event, is not appropriate to describe the data set in its entirety. Instead, we find that already the lowest experimental exciton density requires a multi-exciton description, where according to our MC simulations

~30% of the excitons are involved in at least two annihilation events (Supplementary Note 6.6). Evidence for these processes is encoded in even higher-order (i.e., at least seventh-order) 2D spectra, which have indeed been observed experimentally (Supplementary Note 7).

Now we are in position to elucidate the changes of the exciton dynamics induced by the presence of the outer layer, which involve both intra- and inter-tube exciton interactions. In analogy with the isolated inner tubes, the diagonal peaks in the EEI2D spectra for the inner and outer tube reveal annihilation of excitons that were initially planted on the same layer (Supplementary Figure 17). The salient differences of the dynamics of the complete nanotubes compared to the isolated inner tubes arise from the inter-tube exciton transfer (ET), which is evident from the mere existence of the cross peaks in the absorptive and EEI2D spectra (Figure 2b). These peaks reveal coupling of the individual layers, which leads to an inter-layer exchange of excitons on a sub-ps timescale. Hence, the additional information on specific exciton trajectories including inter-layer ET and EEA is encoded in the absorptive and EEI cross peaks, whose maxima are found to gradually shift to earlier waiting times for increasing exciton densities (Figure 4a), while their amplitudes saturate for the highest exciton density similarly to the trend found for the inner tubes.

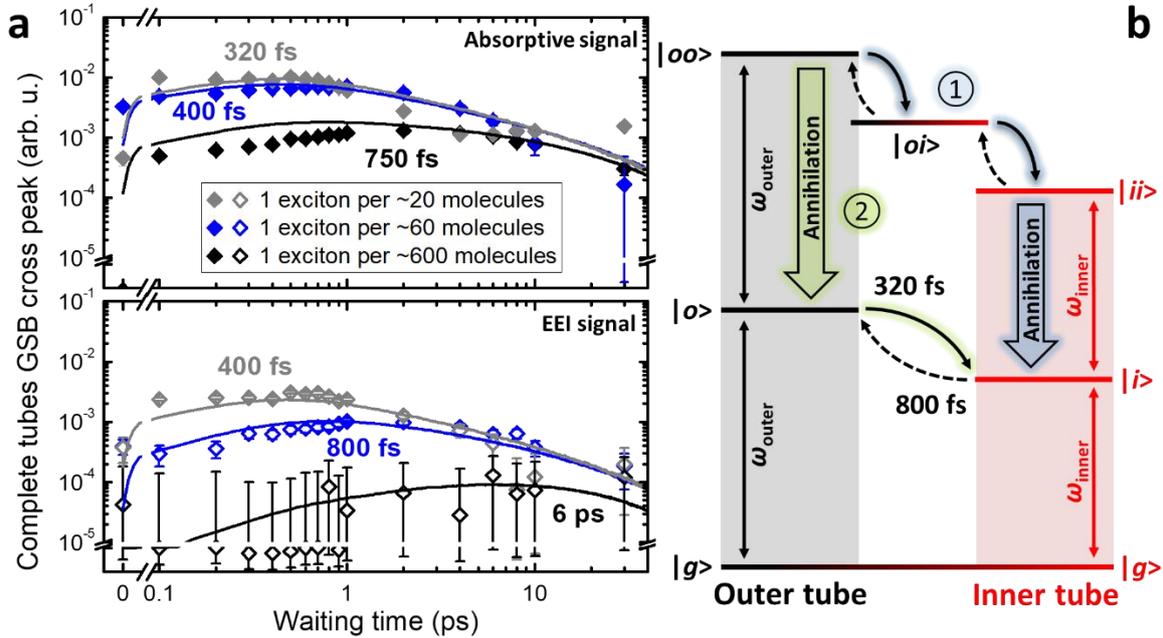

**Figure 4. Absorptive and EEI cross peak transients with corresponding level diagram.** (**a**) Log-log plot of the absorptive (upper panel, solid diamonds) and EEI (lower panel, open diamonds) GSB/SE transients for the cross peak between outer and inner layer at different exciton densities. The transients were obtained by integrating the signal in the rectangular regions of interest shown in Figure 2b. The absorptive cross peak maps ET from the outer to the inner tube ($\omega_{outer} \to \omega_{inner}$), while the EEI cross-peak maps the subsequent occurrence of EEA and ET of two excitons from the outer tube ($2\omega_{outer} \to \omega_{inner}$). The amplitude (vertical) scaling is identical to those in Figures 3 and 4. The error bars refer to the detection noise level in the experiment, i.e., the standard error of the background fluctuations in the respective spectral region during each measurement (Supplementary Note 2). The solid lines depict the results from MC simulations of the exciton dynamics with parameters summarized in Table 1. For each fitting curve the delay time at which the maximum signal occurs is explicitly stated. (**b**) Energy level diagram of the double-walled nanotubes illustrating bi-exciton (annihilation) pathways 1 (blue) and 2 (green) in presence of both tubes. Optical transitions of the inner and outer tube are marked by vertical arrows and their corresponding frequencies. Curved solid (dashed) arrows depict downhill (uphill) ET pathways with their time constants indicated.

Dissecting the individual contributions to the EEI cross peak is crucial to unravel the effect of the multi-layered structure for the observed exciton dynamics, yet intrinsically challenging due to the wealth of possible exciton trajectories. Therefore, we limit our analysis to the EEI cross peak linking the creation of two excitons on the outer layer with the detection of a single exciton on the inner layer, i.e., $2\omega_{outer} \rightarrow \omega_{inner}$ (see Supplementary Note 5.2 for the corresponding Feynman diagrams). We consider this process dominant for two reasons: first, the total (initial) number of excitons on the outer tube is significantly larger as its absorption cross-section is a factor of ~2 higher than for the inner tubes and, second, at early waiting times the majority of ET events occurs from the outer to the inner tube (i.e., downhill in energy). We extend the three-level system of the isolated inner tubes by also including the one- and bi-excitonic states of the outer tube as $|o\rangle$ and $|oo\rangle$ (Figure 4b). We assume that EEA can only occur from bi-excitonic states populating the same tube ($|oo\rangle$ and $|ii\rangle$) and not from the mixed population state $|oi\rangle$, which describes two single excitons residing on spatially separated domains on each tube. This assumption is based on the fact that due to the wall separation of ~3.5 nm the inter-tube dipole-dipole interactions that are responsible for EEA are negligibly small compared to the dipole-dipole interactions within the same tube[9,10]. Nevertheless, we consider the mixed state as one of the pathways via which excitons from the outer tube bi-excitonic state can be transferred to the inner tube bi-excitonic state prior to any EEA.

At zero waiting time, neither an absorptive nor an EEI cross peak is expected, since excitons have no time to undergo ET and EEA. For finite waiting times, however, the EEI cross peak is dominated by processes that simultaneously include EEA and ET. EEA can occur via two annihilation channels: (1) ET of two excitons created on the outer tube followed by EEA on the inner tube (Figure 4b; highlighted in blue), or (2) EEA on the outer tube followed by ET of the surviving exciton to the inner tube (Figure 4b; highlighted in green). Whether (1) or (2) is the prevalent annihilation channel is determined by the balance between the ET and EEA rates. Note that the particular order of ET and EEA during the population time is spectroscopically not distinguishable by examining the cross peak dynamics alone. However, in combination with the respective dynamics of the EEI diagonal peaks a conclusive picture of individual exciton trajectories is obtained.

At the lowest exciton density, a delayed emergence of the EEI cross peak with a maximum at ~6 ps is observed (Figure 4a, black). In this regime the EEA rate is significantly lower than the ET rate so that the timescale of signal formation is consistent with the EEI signal of the isolated inner tubes. Taken together with the negligibly small EEI signal of the outer tube at this exciton density (Supplementary Figure 17a, black) this proves that excitons are harvested by the outer tube and rapidly transferred to the inner tube, where they diffuse and eventually decay, either naturally or via EEA. Therefore, the inner tube acts as an exciton accumulator, which behaves in close analogy to natural systems, where excitation transport is directed via spatio-energetic tuning of the corresponding sites[37,40,58].

At intermediate exciton densities, the vast majority of the EEA events occurs on the outer tube, which is evident from a steep rise of the EEI signal of the outer tube (a), while the inner layer accumulates the already-reduced population of the surviving excitons for which EEA is less pronounced. As a result, the EEI cross peak dynamics are reminiscent to those of the (almost) annihilation-free absorptive cross peak due to balancing of the ET and EEA rates (Figure 4a, blue and Supplementary Note 6.7).

For the highest exciton density, the EEA rate exceeds the ET rate. Consequently, the exciton population of the outer tube becomes strongly depleted by EEA prior to any ET. Simultaneously, a significant share of the excitons is transferred to the inner tube resulting in the emergence of the EEI cross peak for which the bottleneck of the rise time is given by the ET rate. In addition, the occurrence of multi-exciton processes gains significance and further reduces the exciton population of the outer tube beyond the two-exciton annihilation picture (Supplementary Note 6.7 and 7), which drastically lowers the fraction of excitons that could be transferred to the inner tube. As a result, the EEI cross peak maximum further shifts towards earlier waiting times (Figure 4a, gray), while the amplitude of both absorptive and EEI cross peaks saturates thereby indicating the loss of excitons and, thus, a lower number of transfer events. In the limiting case of instantaneous annihilation of all excitons residing on the outer tube, the formation of the cross peak would be entirely inhibited. In such a way, the inner tube is protected at high excitation fluences from overburning with excitons transferred from the outer tube by the rapid annihilation of excitons on the outer tube prior to any transfer.

In order to analyze the observed exciton dynamics, we extend the MC simulations to the case of complete nanotubes. A second layer was added to the molecular grid to represent the outer tube in which the grid size is larger than that of the inner layer in accordance with the increased diameter of the outer tube. The exciton density for the inner and outer tube was set identical (Supplementary Note 1). The excitons are allowed to switch between the adjacent (unoccupied) molecules on the inner and outer layer at the rates specified in Supplementary Table 4. Otherwise all parameters are kept identical from the simulations of the isolated inner tubes except the one-exciton lifetime that was measured as 33 ps (Supplementary Note 10). We extract the absorptive and EEI signals from the MC simulations by evaluating the number of excitons that meet a certain set of prerequisites (Supplementary Table 3). For example, the EEI cross peak ($2\omega_{outer} \rightarrow \omega_{inner}$) is computed as the number of excitons that have been (1) originally planted on the outer tube, (2) participated in at least one annihilation event with an exciton from the same tube, and (3) reside on the inner tube at time $T$. We find excellent agreement between experimental data (symbols) and simulations (solid lines) in Figure 4a and Supplementary Figure 17 by applying the same model parameters for the exciton diffusion and annihilation radius as for the isolated inner tube with exception of the inter-layer ET.

In order to test the exciton diffusion result obtained from our experiments and MC simulation, we also calculated the exciton diffusion constant tensor of C8S3 nanotubes using an extended version of the Haken-Strobl-Reineker model[16,59–61]; see Methods section and Supplementary Note 11. From the calculation, we obtained the diffusion constant along the axial direction equal to 23.9 nm$^2$ ps$^{-1}$ for the inner wall and 16.3 nm$^2$ ps$^{-1}$ for the outer wall of the C8S3 double-walled tube (Supplementary Note 11.2). Taken together with a surface density of 1.8 molecules nm$^{-2}$, where each site contains a unit cell with two molecules, this translates into 43 and 29 molecules ps$^{-1}$ for the inner and outer wall, respectively. These values agree reasonably well with the results obtained from combined experiment and MC simulations of 10 molecules ps$^{-1}$ for both tubes, considering the simplicity of the underlying model for the MC simulations.

Previous measurements of the exciton diffusion constants of supramolecular nanostructures revealed typical values on the order of 100 nm$^2$ ps$^{-1}$ at room temperature assuming purely one-dimensional exciton diffusion[11,17,62], although higher values up to 300-600 nm$^2$ ps$^{-1}$ and even 5500 nm$^2$ ps$^{-1}$ have also been reported[13,63]. These diffusion constants are usually estimated to fall between the limiting cases of fully coherent and purely diffusive transport and, thus, should be considered as an effective diffusion constant with contributions from both processes. Note that it was not possible to obtain a good fit of the experimental data for a purely diffusive model with the diffusion constant increased to 100 nm$^2$ ps$^{-1}$ (Supplementary Note 12).

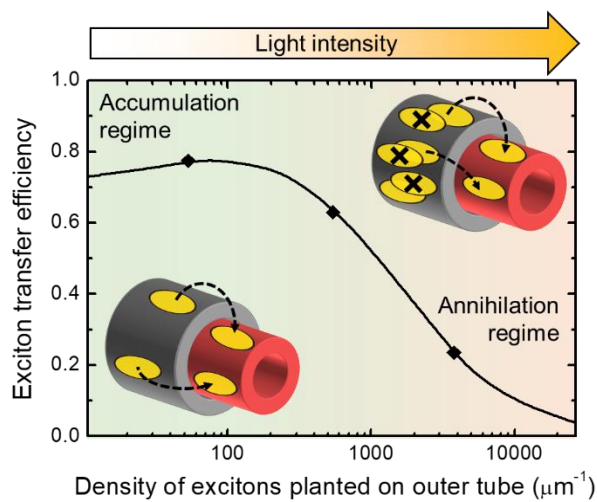

**Figure 5. Exciton transfer regimes.** Exciton transfer efficiency, i.e., fraction of excitons that were planted on the outer tube and either decayed naturally or annihilated on the inner tube as a function of linear exciton density (i.e., the number of excitons per unit of nanotube length), obtained from MC

simulations (black line). Symbols indicate exciton densities used in the experiments. In the simulations also the inner tube is populated with excitons at time zero with the same exciton density as the outer tube. The insets schematically depict the exciton dynamics in the accumulation regime (bottom left) and the annihilation regime (top right). Dashed arrows: exciton transfer; black crosses: exciton–exciton annihilation.

Figure 5 summarizes the main findings of this work as a plot of exciton transfer efficiency *versus* exciton density. At low exciton densities, the transfer efficiency converges to the value of ~0.7, which is determined by the condition that the exciton populations residing on the inner and outer tube eventually reach thermal equilibrium[64,65]; Supplementary Note 6.4. At high exciton densities, the dynamics are dominated by exciton–exciton annihilation on the outer tube, which substantially reduces the fraction of transferred excitons and, thus, leads to a reduced transfer efficiency. The maximum indicates optimal balancing between a low degree of exciton–exciton annihilation on the outer layer, fast inter-layer exciton transfer and subsequent annihilation of the transferred excitons on the inner layer (Supplementary Note 6.7).

Finally, we comment briefly on the effect of exciton delocalization on the exciton–exciton annihilation process. Like exciton transport, exciton–exciton annihilation can either proceed in a hopping Förster-like mechanism[53,57,66] or in a wavelike fashion[67]. While the exciton transport is determined by the energies and couplings of the ground-state transitions of individual molecules that also lead to exciton delocalization, exciton–exciton annihilation involves coupling through higher excited states[68]. Consequently, the phenomena of exciton delocalization and exciton–exciton annihilation are closely related, but their relationship is not straightforward. The here presented combination of higher-order nonlinear spectroscopy and controlled structural complexity has the potential to unravel the connection between exciton transport (be it wavelike or diffusive) and exciton–exciton annihilation. Clearly, more theoretical support is needed to fully disentangle these processes, as the annihilation may also depend in a non-trivial way on the phases of the wavefunctions of the involved excitons[69].

*Conclusions*

In conclusion, we have unambiguously identified the excitonic properties of a complex supramolecular system by utilizing a novel spectroscopic microfluidic approach. Microfluidic flash-dilution allowed manipulating the structural hierarchy of the supramolecular system on the nanoscale via controlled destruction of individual sub-units of the assembly. This provided a direct view on the simplified structure whose spectral response would otherwise have been concealed due to congested spectroscopic features. Assignment of the excitonic properties was performed by employing exciton–exciton-interaction two-dimensional (EEI2D) spectroscopy, which is capable of isolating mutual interactions of individual excitons. Application of this technique to double-walled nanotubes together with extensive theoretical modelling allowed retrieving a unified set of excitonic properties for the exciton diffusion and exciton–exciton interactions for both layers.

In the arrangement of the double-wall nanotubes, the outer layer appears to act as an exciton antenna, which under strong excitation fluences leads to fast EEA rates prior to any inter-layer ET. At low exciton densities, the inner tube acts as an exciton accumulator absorbing the majority of the excitons from the outer layer. In this capacity, our findings shed light on the importance of the multi-layered, hierarchical structure for the functionality of the light-harvesting apparatus in which the already beneficial excitonic properties of individual sub-units are retained in a more complex double-walled assembly. Hence, the excitonic properties of the supramolecular assembly can be considered robust against variations in the inter-layer transport despite the weak electronic coupling between the layers and the lack of inter-layer exciton coherences. Such excitonic robustness paired with fast inter-layer exciton transfer would prove key for efficient exciton transfer in natural chlorosomes due to close similarity of their telescopic structure with the double-wall nanotubes considered herein. Moreover, we envision that the versatility of the microfluidic approach paired with higher-order 2D spectroscopy opens the door to further expedite a better fundamental understanding of the excitonic properties of supramolecular assemblies and, thereby, will encompass rational design principles for future applications of such materials in opto-electronic devices.

*Methods*

*Materials and sample preparation.* C8S3 nanotubes were prepared via the alcoholic route as described elsewhere[10]. The aggregation of the dye molecule 3,3'-bis(2-sulfopropyl)-5,5',6,6-tetrachloro-1,1'-dioctylbenzimidacarbocyanine (C8S3, M = 903 g mol$^{-1}$) purchased from FEW Chemicals GmbH (Wolfen, Germany) into double-walled nanotubes was verified by linear absorption spectroscopy prior to any other experiments. In order to minimize the thermodynamically induced formation of thicker bundles of nanotubes, sample solutions were freshly prepared for every experiment and used within three days.

*Steady-state absorption.* Steady-state absorption spectra were recorded using either a PerkinElmer Lambda 900 UV/VIS/NIR or a Jasco V-670 UV-Vis spectrometer. The sample solution was put either in a 200 µm cuvette (Hellma Analytics, Germany) or a 1 mm quartz cuvette (Starna GmbH, Germany). For the latter case, the sample solutions were diluted with Milli-Q water by a dilution factor between 2 and 3.5.

*Microfluidic flash-dilution.* Microfluidic flash-dilution of C8S3 nanotubes was achieved in a tear-drop mixer (micronit, the Netherlands) by mixing neat sample solution with a diluting agent (50:50 mixture of water and methanol by volume) at a flowrate ratio of 5:7. Measurements on the complete nanotubes were conducted by replacing the diluting agent (water and methanol) with Milli-Q water, which only dilutes the sample and does not induce flash-dilution of the outer layer. All solutions were supplied by syringe pumps (New Era, model NE-300). For EEI2D experiments the mixed sample solution was relayed to a transparent thin-bottom microfluidic flow-cell (micronit, the Netherlands) with a channel thickness of 50 µm and a width of 1 mm. With these parameters a maximum optical density of 0.1–0.2 was reached.

*Exciton–Exciton Interaction 2D (EEI2D) spectroscopy.* More details on the experimental setup are published elsewhere[45]; a schematic of the setup is shown in Supplementary Figure 23. In brief, the output of a Ti:Sapphire-Laser (Spitfire Pro, Spectra Physics, 1 kHz repetition rate) was focused into a fused-silica hollow-core fiber (UltraFast Innovations) filled with Argon to generate a broadband white-light continuum. The main fraction of the light was used as the pump beam and guided through a grism compressor and for further compression through an acousto-optical programmable dispersive filter (DAZZLER, Fastlite, France) to achieve a pulse width of ~15 fs at the sample position (verified via SHG-FROG measurements). The DAZZLER was also used for spectral selection of the excitation spectrum. The remaining fraction of the white-light continuum was used as the probe beam and delayed relative to the pump beam by passing a motorized delay stage (M-IMS600LM, Newport). Both beams were then focused and spatially overlapped in a microfluidic channel under a small angle of 2°. The intensity FWHM of the pump and probe focal spots at the sample position were ~140 µm and ~80 µm, respectively, to minimize the intensity variation of the pump beam over the profile of the probe beam. The polarization of both beams was set parallel to the flow direction of the sample. After passing the sample the spectrum of the probe beam was measured by a CCD camera.

In order to measure 2D spectra the DAZZLER was used to split the pump pulses into two phase-locked time-delayed replica, the delay between which was scanned from 0 fs to 197.6 fs in steps of 0.38 fs. This choice set the resolution along the excitation axis and the Nyquist limit to 84 cm$^{-1}$ and 44000 cm$^{-1}$, respectively. The resolution of the probe axis (20 cm$^{-1}$) was fixed by the detector (ActonSpectraPro 2558i and Pixis 2 K camera, Princeton Instruments). In order to isolate the desired 2D signal from unwanted contributions due to background and scattering, the pump and the probe beams were both synchronously modulated by two choppers (MC2000, Thorlabs). All four possible combinations were measured: both beams open, only probe open, only pump open, and both beams blocked. Each contribution was averaged over 5 consecutive laser pulses by modulating the pump and probe beam at 200 Hz and 100 Hz, respectively. In order to ensure that the spectral region of interest is free of any artifacts from the experimental apparatus, blank experiments were performed on an annihilation-free sample (sulforhodamine 101 dissolved in water; Supplementary Note 14).

The different data sets of the double-walled nanotubes were measured at pulse energies of the pump pulse of 20, 5, and 0.5 nJ corresponding to exciton densitites of 19 ± 7, 64 ± 23, and 625 ± 228 monomeric units per exciton (Supplementary Note 1). The uncertainty of the exciton density was computed via propagation of uncertainty of all relevant input parameters. For the flash-diluted samples pulse energies of 20, 5, 2.5, and 1 nJ were used corresponding to 18 ± 8, 83 ± 38, 165 ± 75, and 404 ± 185 monomeric units per exciton. The pulse energies were measured at zero time delay of the double pulse.

*Monte-Carlo (MC) simulations.* MC simulations of the exciton populations were performed for isolated inner tubes and complete nanotubes represented by a single and two coupled planes, respectively. Each plane comprised a square grid of molecules with periodic boundary conditions in either direction. The length of the planes was set to 1000 molecules, while the lateral grid size was chosen as 55 molecules (outer tube) and 30 molecules (inner tube) and a lattice constant of 0.74 nm as derived from previously published theoretical models (Ref. [9] and Supplementary Note 6.1). For isolated inner tubes, only the inner plane was used. At time zero, excitons were randomly planted on the molecular grid according to the experimental exciton density. Thereafter, the excitons performed a 2D random walk on the grid (with a hopping probability $H$ to move to any of the neighbouring molecules) with a timestep of 1 fs. In addition, at each step they could be transferred between adjacent molecules on the inner/outer layer or undergo exciton–exciton annihilation causing the instant deletion of one of the excitons. The latter occurred with probability of one under the condition that two excitons approach each other closer than the annihilation radius (Supplementary Note 6.3). Excitons were not constrained from (sequential) participation in multiple annihilation events, for which experimental evidence is provided by the observation of higher order signals (Supplementary Note 5.3 and 7). No anisotropic exciton transport (Supplementary Note 11) was included in the MC simulations, but instead the hopping rates were set identical for inner and outer tube in all directions.

In order to extract the absorptive and EEI signals from the MC simulations, all excitons were labelled with their zero-time position as well as their participation in an annihilation event with an exciton that was originally planted on the same tube. At each time step of the MC simulation the number of excitons was evaluated that met a certain set of prerequisites (Supplementary Table 3). Taking only exciton populations into account (i.e., diagonal entries in a density-matrix description) neglects any possible exciton coherences in the system, which we justify with previously reported findings that any coherence in this system does not survive longer than a few hundred fs[70] and the absence of coherent beatings in the cross peak signal from conventional 2D spectroscopy (Supplementary Note 15). For comparison with the experimental results, the simulation transients for the absorptive signals were scaled with identical coefficients to obtain the best fit with experimental data; the same was done for the EEI signals.

*Haken-Strobl-Reineker model.* In order to calculate the exciton diffusion tensor of C8S3 nanotubes, we adopted the same molecular structure for the nanotubes as reported by Eisele *et al.*[9] The individual tensor elements were then calculated using the following equation:

$$D_{\vec{u},\vec{w}} = \frac{1}{Z} \sum_{\mu,\nu=1}^{N} \frac{\Gamma}{\Gamma^2 + (\omega_{\mu\nu})^2} \hat{j}_{\mu\nu}^*(\vec{u}) \hat{j}_{\mu\nu}(\vec{w}) \exp\left(\frac{-\hbar\omega_\nu}{k_B T}\right).$$

Here, $\mu$ and $\nu$ run over all the $N$ collective exciton states, obtained by diagonalizing the exciton Hamiltonian for the tube considered (Ref. [9] and Supplementary Note 11.1), $\Gamma$ is the dephasing rate that characterizes the Haken-Strobl-Reineker model of white noise thermal fluctuations[16,59–61] and $\hbar\omega_{\mu\nu} = \hbar(\omega_\mu - \omega_\nu)$ is the energy difference between exciton states $\mu$ and $\nu$. Furthermore, $\hat{j}_{\mu\nu}(\vec{u}) = i \sum_{n,m=1}^{N} \langle \mu|m\rangle (\vec{u} \cdot \vec{r}_{mn}) J_{nm} \langle n|\nu\rangle$ is the flux operator along direction $\vec{u}$ in the exciton eigenstate basis, where $n, m$ run over all the molecules in the aggregate, $\vec{r}_{mn} = \vec{r}_m - \vec{r}_n$ is the relative separation vector between molecules $m$ and $n$, and $J_{nm}$ is the excitation transfer (dipole-dipole) interaction between them. The Boltzmann factor $\exp\left(\frac{-\hbar\omega_\nu}{k_B T}\right)$ is used to account in a simple way for a temperature $T$ smaller than the exciton bandwidth and $Z = \sum_{\nu=1}^{N} \exp\left(\frac{-\hbar\omega_\nu}{k_B T}\right)$ is the exciton partition function. An asterisk (*) on $\hat{j}_{\mu\nu}(\vec{u})$ refers to complex conjugation of the operator.

A detailed derivation of the above equation excluding the Boltzmann factor can be found elsewhere[16]. For the C8S3 nanotubes, each wall has a diffusion tensor, characterized by the tensor elements $D_{z,z}$, $D_{z,\phi}$, $D_{\phi,z}$ and $D_{\phi,\phi}$, where $z$ is the axial direction and $\phi$ is the direction along the circumference of the tube. Further details are given in Supplementary Note 11.

*Acknowledgments*

B.K. and M.S.P. gratefully acknowledge numerous discussions with A. S. Bondarenko, I. Patmanidis, A. H. de Vries, and S. J. Marrink. F. de Haan is greatly acknowledged for writing the code for Monte-Carlo simulations as well as general laboratory assistance.

*Financial Support*

B.K. and M.S.P. acknowledge funding by the Dieptestrategie Programme of the Zernike Institute for Advanced Materials (University of Groningen, the Netherlands). M.S.P. has also received funding from the European Union's Horizon2020 research and innovation programme under Marie Sklodowska Curie Grant No. 722651. T.B. acknowledges funding by the European Research Council (ERC) – Grant No. 614623, the Deutsche Forschungsgemeinschaft (DFG, German Research Foundation) – Grant No. BR 2123/13-1, and the Bavarian State Ministry of Science, Research, and the Arts – Collaborative Research Network "Solar Technologies Go Hybrid".

*Author contributions*

B.K. prepared the samples. B.K. and J.L. performed the absorptive and EEI2D experiments and analyzed the experimental data, together with P.M.; the analysis was supervised by T.B. and M.S.P. B.K. performed the Monte-Carlo simulations. T.K. calculated the exciton diffusion tensor under supervision of T.L.C.J. and J.K. J.K. led the discussion on the link between exciton delocalisation and annihilation. B.K. and M.S.P. wrote the manuscript with contributions from all other authors.

*Additional Information*

The authors declare no competing financial interests.

Supplementary information is available.

**References**

1. Mirkovic, T.; Ostroumov, E. E.; Anna, J. M.; van Grondelle, R.; Govindjee; Scholes, G. D. Light Absorption and Energy Transfer in the Antenna Complexes of Photosynthetic Organisms. *Chem. Rev.* **117,** 249–293 (2017).

2. Orf, G. S.; Blankenship, R. E. Chlorosome antenna complexes from green photosynthetic bacteria. *Photosynth. Res.* **116,** 315–331 (2013).

3. Brixner, T.; Hildner, R.; Köhler, J.; Lambert, C.; Würthner, F. Exciton Transport in Molecular Aggregates - From Natural Antennas to Synthetic Chromophore Systems. *Adv. Energy Mater.* **7,** 1700236 (2017).

4. Pšenčík, J.; Ikonen, T. P.; Laurinmäki, P.; Merckel, M. C.; Butcher, S. J.; Serimaa, R. E.; Tuma, R. Lamellar Organization of Pigments in Chlorosomes, the Light Harvesting Complexes of Green Photosynthetic Bacteria. *Biophys. J.* **87,** 1165–1172 (2004).

5. Ganapathy, S.; Oostergetel, G. T.; Wawrzyniak, P. K.; Reus, M.; Chew, A. G. M.; Buda, F.; Boekema, E. J.; Bryant, D. A.; Holzwarth, A. R.; de Groot, H. J. M. Alternating syn-anti bacteriochlorophylls form concentric helical nanotubes in chlorosomes. *Proc. Natl. Acad. Sci.* **106,** 8525–8530 (2009).

6. Günther, L. M.; Jendrny, M.; Bloemsma, E. A.; Tank, M.; Oostergetel, G. T.; Bryant, D. A.; Knoester, J.; Köhler, J. Structure of Light-Harvesting Aggregates in Individual Chlorosomes. *J. Phys. Chem. B* **120,** 5367–5376 (2016).

7. Günther, L. M.; Löhner, A.; Reiher, C.; Kunsel, T.; Jansen, T. L. C.; Tank, M.; Bryant, D. A.; Knoester, J.; Köhler, J. Structural Variations in Chlorosomes from Wild-Type and a bchQR

Mutant of Chlorobaculum tepidum Revealed by Single-Molecule Spectroscopy. *J. Phys. Chem. B* **122,** 6712–6723 (2018).

8. Würthner, F.; Kaiser, T. E.; Saha-Möller, C. R. J-Aggregates: From Serendipitous Discovery to Supramolecular Engineering of Functional Dye Materials. *Angew. Chemie Int. Ed.* **50,** 3376–3410 (2011).

9. Eisele, D. M.; Cone, C. W.; Bloemsma, E. A.; Vlaming, S. M.; van der Kwaak, C. G. F.; Silbey, R. J.; Bawendi, M. G.; Knoester, J.; Rabe, J. P.; Vanden Bout, D. A. Utilizing redox-chemistry to elucidate the nature of exciton transitions in supramolecular dye nanotubes. *Nat. Chem.* **4,** 655–662 (2012).

10. Kriete, B.; Bondarenko, A. S.; Jumde, V. R.; Franken, L. E.; Minnaard, A. J.; Jansen, T. L. C.; Knoester, J.; Pshenichnikov, M. S. Steering Self-Assembly of Amphiphilic Molecular Nanostructures via Halogen Exchange. *J. Phys. Chem. Lett.* **8,** 2895–2901 (2017).

11. Clark, K. A.; Krueger, E. L.; Vanden Bout, D. A. Direct measurement of energy migration in supramolecular carbocyanine dye nanotubes. *J. Phys. Chem. Lett.* **5,** 2274–2282 (2014).

12. Haedler, A. T.; Kreger, K.; Issac, A.; Wittmann, B.; Kivala, M.; Hammer, N.; Köhler, J.; Schmidt, H. W.; Hildner, R. Long-range energy transport in single supramolecular nanofibres at room temperature. *Nature* **523,** 196–199 (2015).

13. Caram, J. R.; Doria, S.; Eisele, D. M.; Freyria, F. S.; Sinclair, T. S.; Rebentrost, P.; Lloyd, S.; Bawendi, M. G. Room-Temperature Micron-Scale Exciton Migration in a Stabilized Emissive Molecular Aggregate. *Nano Lett.* **16,** 6808–6815 (2016).

14. Jin, X.-H.; Price, M. B.; Finnegan, J. R.; Boott, C. E.; Richter, J. M.; Rao, A.; Menke, S. M.; Friend, R. H.; Whittell, G. R.; Manners, I. Long-range exciton transport in conjugated polymer nanofibers prepared by seeded growth. *Science* **360,** 897–900 (2018).

15. Eisele, D. M.; Arias, D. H.; Fu, X.; Bloemsma, E. A.; Steiner, C. P.; Jensen, R. A.; Rebentrost, P.; Eisele, H.; Tokmakoff, A.; Lloyd, S.; *et al.* Robust excitons inhabit soft supramolecular nanotubes. *Proc. Natl. Acad. Sci.* **111,** E3367–E3375 (2014).

16. Chuang, C.; Lee, C. K.; Moix, J. M.; Knoester, J.; Cao, J. Quantum Diffusion on Molecular Tubes: Universal Scaling of the 1D to 2D Transition. *Phys. Rev. Lett.* **116,** 196803 (2016).

17. Kim, T.; Ham, S.; Lee, S. H.; Hong, Y.; Kim, D. Enhancement of exciton transport in porphyrin aggregate nanostructures by controlling the hierarchical self-assembly. *Nanoscale* **10,** 16438–16446 (2018).

18. Kirstein, S.; von Berlepsch, H.; Böttcher, C. Photo-induced reduction of Noble metal ions to metal nanoparticles on tubular J-aggregates. *Int. J. Photoenergy* **2006,** 1–7 (2007).

19. Clark, K. A.; Cone, C. W.; Vanden Bout, D. A. Quantifying the Polarization of Exciton Transitions in Double-Walled Nanotubular J-Aggregates. *J. Phys. Chem. C* **117,** 26473–26481 (2013).

20. Pandya, R.; Chen, R.; Cheminal, A.; Thomas, T. H.; Thampi, A.; Tanoh, A.; Richter, J. M.; Shivanna, R.; Deschler, F.; Schnedermann, C.; *et al.* Observation of Vibronic Coupling Mediated Energy Transfer in Light-Harvesting Nanotubes Stabilized in a Solid-State Matrix. *J. Phys. Chem. Lett.* **9,** 5604–5611 (2018).

21. Whitesides, G. M. The origins and the future of microfluidics. *Nature* **442,** 368–373 (2006).

22. DeMello, A. J. Control and detection of chemical reactions in microfluidic systems. *Nature* **442,** 394–402 (2006).

23. Arnon, Z. A.; Vitalis, A.; Levin, A.; Michaels, T. C. T.; Caflisch, A.; Knowles, T. P. J.; Adler-Abramovich, L.; Gazit, E.; Lehn, J.-M.; Brunsveld, L.; *et al.* Dynamic microfluidic control of supramolecular peptide self-assembly. *Nat. Commun.* **7,** 13190 (2016).

24. Sorrenti, A.; Rodriguez-Trujillo, R.; Amabilino, D. B.; Puigmartí-Luis, J. Milliseconds make the difference in the far-from-equilibrium self-assembly of supramolecular chiral nanostructures. *J. Am. Chem. Soc.* **138,** 6920–6923 (2016).

25. Chan, K. L. A.; Kazarian, S. G. FT-IR spectroscopic imaging of reactions in multiphase flow in microfluidic channels. *Anal. Chem.* **84,** 4052–4056 (2012).

26. Benninger, R. K. P.; Hofmann, O.; McGinty, J.; Requejo-Isidro, J.; Munro, I.; Neil, M. A. A.; deMello, A. J.; French, P. M. W. Time-resolved fluorescence imaging of solvent interactions in microfluidic devices. *Opt. Express* **13,** 6275 (2005).

27. Batabyal, S.; Rakshit, S.; Kar, S.; Pal, S. K. An improved microfluidics approach for monitoring real-time interaction profiles of ultrafast molecular recognition. *Rev. Sci. Instrum.* **83,** 043113 (2012).

28. Maillot, S.; Carvalho, A.; Vola, J.-P.; Boudier, C.; Mély, Y.; Haacke, S.; Léonard, J. Out-of-equilibrium biomolecular interactions monitored by picosecond fluorescence in microfluidic droplets. *Lab Chip* **14,** 1767 (2014).

29. Chauvet, A.; Tibiletti, T.; Caffarri, S.; Chergui, M. A microfluidic flow-cell for the study of the ultrafast dynamics of biological systems. *Rev. Sci. Instrum.* **85,** 103118 (2014).

30. Tracy, K. M.; Barich, M. V.; Carver, C. L.; Luther, B. M.; Krummel, A. T. High-Throughput Two-Dimensional Infrared (2D IR) Spectroscopy Achieved by Interfacing Microfluidic Technology with a High Repetition Rate 2D IR Spectrometer. *J. Phys. Chem. Lett.* **7,** 4865–

4870 (2016).

31. Brüggemann, B.; Pullerits, T. Nonperturbative modeling of fifth-order coherent multidimensional spectroscopy in light harvesting antennas. *New J. Phys.* **13,** 025024 (2011).

32. Olbrich, C.; Jansen, T. L. C.; Liebers, J.; Aghtar, M.; Strümpfer, J.; Schulten, K.; Knoester, J.; Kleinekathöfer, U. From Atomistic Modeling to Excitation Transfer and Two-Dimensional Spectra of the FMO Light-Harvesting Complex. *J. Phys. Chem. B* **115,** 8609–8621 (2011).

33. Dostál, J.; Benešová, B.; Brixner, T. Two-dimensional electronic spectroscopy can fully characterize the population transfer in molecular systems. *J. Chem. Phys.* **145,** 124312 (2016).

34. Malý, P.; Mančal, T. Signatures of Exciton Delocalization and Exciton–Exciton Annihilation in Fluorescence-Detected Two-Dimensional Coherent Spectroscopy. *J. Phys. Chem. Lett.* **9,** 5654–5659 (2018).

35. Süß, J.; Wehner, J.; Dostál, J.; Brixner, T.; Engel, V. Mapping of exciton–exciton annihilation in a molecular dimer via fifth-order femtosecond two-dimensional spectroscopy. *J. Chem. Phys.* **150,** 104304 (2019).

36. Kunsel, T.; Tiwari, V.; Matutes, Y. A.; Gardiner, A. T.; Cogdell, R. J.; Ogilvie, J. P.; Jansen, T. L. C. Simulating Fluorescence-Detected Two-Dimensional Electronic Spectroscopy of Multichromophoric Systems. *J. Phys. Chem. B* **123,** 394–406 (2019).

37. Brixner, T.; Stenger, J.; Vaswani, H. M.; Cho, M.; Blankenship, R. E.; Fleming, G. R. Two-dimensional spectroscopy of electronic couplings in photosynthesis. *Nature* **434,** 625–628 (2005).

38. Abramavicius, D.; Nemeth, A.; Milota, F.; Sperling, J.; Mukamel, S.; Kauffmann, H. F. Weak

exciton scattering in molecular nanotubes revealed by double-quantum two-dimensional electronic spectroscopy. *Phys. Rev. Lett.* **108,** 67401 (2012).

39. Lim, J.; Palecek, D.; Caycedo-soler, F.; Lincoln, C. N.; Prior, J.; Berlepsch, H. Von; Paleček, D.; Caycedo-soler, F.; Lincoln, C. N.; Prior, J.; *et al.* Vibronic origin of long-lived coherence in an artificial molecular light harvester. *Nat. Commun.* **6,** 1–22 (2015).

40. Dostál, J.; Pšenčík, J.; Zigmantas, D. In situ mapping of the energy flow through the entire photosynthetic apparatus. *Nat. Chem.* **8,** 705–710 (2016).

41. Thyrhaug, E.; Tempelaar, R.; Alcocer, M. J. P.; Žídek, K.; Bína, D.; Knoester, J.; Jansen, T. L. C.; Zigmantas, D. Identification and characterization of diverse coherences in the Fenna–Matthews–Olson complex. *Nat. Chem.* **10,** 780–786 (2018).

42. Tiwari, V.; Matutes, Y. A.; Gardiner, A. T.; Jansen, T. L. C.; Cogdell, R. J.; Ogilvie, J. P. Spatially-resolved fluorescence-detected two-dimensional electronic spectroscopy probes varying excitonic structure in photosynthetic bacteria. *Nat. Commun.* **9,** 4219 (2018).

43. Mattson, M. A.; Green, T. D.; Lake, P. T.; McCullagh, M.; Krummel, A. T. Elucidating Structural Evolution of Perylene Diimide Aggregates Using Vibrational Spectroscopy and Molecular Dynamics Simulations. *J. Phys. Chem. B* **122,** 4891–4900 (2018).

44. Mandal, A.; Chen, M.; Foszcz, E. D.; Schultz, J. D.; Kearns, N. M.; Young, R. M.; Zanni, M. T.; Wasielewski, M. R. Two-Dimensional Electronic Spectroscopy Reveals Excitation Energy-Dependent State Mixing during Singlet Fission in a Terrylenediimide Dimer. *J. Am. Chem. Soc.* **140,** 17907–17914 (2018).

45. Dostál, J.; Fennel, F.; Koch, F.; Herbst, S.; Würthner, F.; Brixner, T. Direct observation of exciton–exciton interactions. *Nat. Commun.* **9,** 2466 (2018).

46. von Berlepsch, H.; Ludwig, K.; Kirstein, S.; Böttcher, C. Mixtures of achiral amphiphilic cyanine dyes form helical tubular J-aggregates. *Chem. Phys.* **385,** 27–34 (2011).

47. Malevich, P.; Heshmatpour, C.; Lincoln, C. N.; Ceymann, H.; Schreck, M. H.; Hauer, J. Ultrafast bi-excitonic dynamics and annihilation in molecular and mesoscopic systems. *EPJ Web Conf.* **205,** 06013 (2019).

48. Hamm, P.; Zanni, M. *Concepts and Methods of 2D Infrared Spectroscopy*. (Cambridge University Press, 2011).

49. Sperling, J.; Nemeth, A.; Hauer, J.; Abramavicius, D.; Mukamel, S.; Kauffmann, H. F.; Milota, F. Excitons and disorder in molecular nanotubes: a 2D electronic spectroscopy study and first comparison to a microscopic model. *J. Phys. Chem. A* **114,** 8179–8189 (2010).

50. Fidder, H.; Knoester, J.; Wiersma, D. A. Observation of the one-exciton to two-exciton transition in a J-aggregate. *J. Chem. Phys.* **98,** 6564 (1993).

51. Bakalis, L. D.; Knoester, J. Pump-probe spectroscopy and the exciton delocalization length in molecular aggregates. *J. Phys. Chem. B* **103,** 6620–6628 (1999).

52. Spano, F. C.; Mukamel, S. Cooperative Nonlinear Optical Response of Molecular Aggregates: Crossover to Bulk Behavior. *Phys. Rev. Lett.* **66,** 1197–1200 (1991).

53. Engel, E.; Leo, K.; Hoffmann, M. Ultrafast relaxation and exciton–exciton annihilation in PTCDA thin films at high excitation densities. *Chem. Phys.* **325,** 170–177 (2006).

54. Valkunas, L.; Ma, Y.-Z.; Fleming, G. R. Exciton-exciton annihilation in single-walled carbon nanotubes. *Phys. Rev. B* **73,** 115432 (2006).

55. Scheblykin, I. G.; Varnavsky, O. P.; Bataiev, M. M.; Sliusarenko, O.; Van der Auweraer, M.;

Vitukhnovsky, A. G. Non-coherent exciton migration in J-aggregates of the dye THIATS: exciton–exciton annihilation and fluorescence depolarization. *Chem. Phys. Lett.* **298,** 341–350 (1998).

56. Lin, H.; Camacho, R.; Tian, Y.; Kaiser, T. E.; Würthner, F.; Scheblykin, I. G. Collective fluorescence blinking in linear J-aggregates assisted by long-distance exciton migration. *Nano Lett.* **10,** 620–626 (2010).

57. Fennel, F.; Lochbrunner, S. Exciton-exciton annihilation in a disordered molecular system by direct and multistep Förster transfer. *Phys. Rev. B* **92,** 140301 (2015).

58. Thyrhaug, E.; Žídek, K.; Dostál, J.; Bína, D.; Zigmantas, D. Exciton Structure and Energy Transfer in the Fenna–Matthews–Olson Complex. *J. Phys. Chem. Lett.* **7,** 1653–1660 (2016).

59. Haken, H.; Reineker, P. The coupled coherent and incoherent exciton motion and its influence on optical absorption, electron spin resonance and nuclear spin resonance. *Acta Univ. Carolinae. Math. Phys.* **14,** 23–47 (1973).

60. Kenkre, V. M.; Reineker, P. *Exciton Dynamics in Molecular Crystals and Aggregates*. (Springer New York, 1982).

61. Moix, J. M.; Khasin, M.; Cao, J. Coherent quantum transport in disordered systems: I. The influence of dephasing on the transport properties and absorption spectra on one-dimensional systems. *New J. Phys.* **15,** 085010 (2013).

62. Eisele, D. M.; Knoester, J.; Kirstein, S.; Rabe, J. P.; Vanden Bout, D. A. Uniform exciton fluorescence from individual molecular nanotubes immobilized on solid substrates. *Nat. Nanotechnol.* **4,** 658–663 (2009).

63. Wan, Y.; Stradomska, A.; Knoester, J.; Huang, L. Direct Imaging of Exciton Transport in Tubular Porphyrin Aggregates by Ultrafast Microscopy. *J. Am. Chem. Soc.* **139,** 7287–7293 (2017).

64. Pugžlys, A.; Augulis, R.; Van Loosdrecht, P. H. M.; Didraga, C.; Malyshev, V. A.; Knoester, J. Temperature-dependent relaxation of excitons in tubular molecular aggregates: fluorescence decay and Stokes shift. *J. Phys. Chem. B* **110,** 20268–20276 (2006).

65. Clark, K. A.; Krueger, E. L.; Vanden Bout, D. A. Temperature-Dependent Exciton Properties of Two Cylindrical J-Aggregates. *J. Phys. Chem. C* **118,** 24325–24334 (2014).

66. Förster, T. Zwischenmolekulare energiewanderung und fluoreszenz. *Ann. Phys.* **437,** 55–75 (1948).

67. Brüggemann, B.; May, V. Exciton exciton annihilation dynamics in chromophore complexes. I. Multiexciton density matrix formulation. *J. Chem. Phys.* **118,** 746–759 (2003).

68. Malyshev, V. A.; Glaeske, H.; Feller, K.-H. Exciton–exciton annihilation in linear molecular aggregates at low temperature. *Chem. Phys. Lett.* **305,** 117–122 (1999).

69. Tempelaar, R.; Jansen, T. L. C.; Knoester, J. Exciton–Exciton Annihilation Is Coherently Suppressed in H-Aggregates, but Not in J-Aggregates. *J. Phys. Chem. Lett.* **8,** 6113–6117 (2017).

70. Yuen-Zhou, J.; Arias, D. H.; Eisele, D. M.; Steiner, C. P.; Krich, J. J.; Bawendi, M. G.; Nelson, K. A.; Aspuru-Guzik, A. Coherent exciton dynamics in supramolecular light-harvesting nanotubes revealed by ultrafast quantum process tomography. *ACS Nano* **8,** 5527–5534 (2014).

# Supporting Information

**Interplay between Structural Hierarchy and Exciton Diffusion in Artificial Light Harvesting**

*Björn Kriete[1], Julian Lüttig[2], Tenzin Kunsel[1], Pavel Malý[2], Thomas L. C. Jansen[1], Jasper Knoester[1], Tobias Brixner[2,3], and Maxim S. Pshenichnikov[1,*]*

[1]University of Groningen, Zernike Institute for Advanced Materials, Nijenborgh 4, 9747 AG Groningen, The Netherlands

[2]Institut für Physikalische und Theoretische Chemie, Universität Würzburg, Am Hubland, 97074 Würzburg, Germany

[3]Center for Nanosystems Chemistry (CNC), Universität Würzburg, Theodor-Boveri-Weg, 97074 Würzburg, Germany

[*] Corresponding author: M.S.Pchenitchnikov@rug.nl

**Table of Contents**

## Supplementary Note 1: Calculation of Exciton Densities

The exciton densities, i.e., the number of excitons ($N_e$) normalized by the number of molecules ($N_m$) in the focal volume, were computed as outlined elsewhere[1]. The formula that was used for computation is given as:

$$\frac{N_e}{N_m} = \frac{\Delta E}{hc_0} \left( \frac{\int I_{\text{pump}}(x,y) I_{\text{probe}}(x,y) dx dy}{\int I_{\text{pump}}(x,y) dx dy \int I_{\text{probe}}(x,y) dx dy} \right) \left( \frac{\int I_{\text{pump}}(\lambda) \lambda \left(1 - 10^{-\text{OD}(\lambda)}\right) d\lambda}{\int I_{\text{pump}}(\lambda) d\lambda} \right) \left( \frac{1}{cN_A d} \right)$$

Here, $\Delta E$ is the pulse energy, $h$ the Planck constant, $c_0$ the speed of light in vacuum and $N_A$ the Avogadro constant. The first bracketed factor describes the spatial overlap of the pump transverse beam profile $I_{\text{pump}}(x,y)$ and the probe beam profile $I_{\text{probe}}(x,y)$ at the sample position, while the second bracketed factor accounts for the spectral overlap of the excitation spectrum $I_{\text{pump}}(\lambda)$ with the sample absorption spectrum at a given optical density $\text{OD}(\lambda)$. Finally, the last bracketed factor counts the number of molecules in the focal volume in the denominator. The latter is proportional to the molar concentration of the sample $c$ and the thickness of the microfluidic channel $d$. The uncertainty of the exciton density was computed via propagation of uncertainty of all relevant input parameters.

In the case of complete nanotubes, the exciton density is considered identical for the inner and outer layer. At a sufficiently low optical density of the sample and assuming similar excitation fluences for both tubes (Figure 1a in the main text), the number of excitons scales with the absorption of the respective tube. The latter in turn scales with the number of molecules in each layer (Supplementary Note 6), which then yields identical exciton densities for the inner and outer tube.

While the calculation of the exciton densities is straightforward in the case of complete nanotubes, special care had to be taken in the case of isolated inner tubes due to the dissolution of the outer wall and, hence, removal of molecules from the experimentally observable spectral window. Because the monomer absorption is strongly blue-shifted with respect to the nanotube absorption ($\lambda_{\text{max}} \approx 520$ nm, Supplementary Figure 1), the second bracketed factor in the above equation already accounts for the reduced spectral overlap. Therefore, only the number of molecules that remains embedded in the inner tube has to be estimated for which we use two different ways, i.e., (1) via the optical density (OD) of the monomer absorption spectrum and (2) directly via the absorption of the inner tubes.

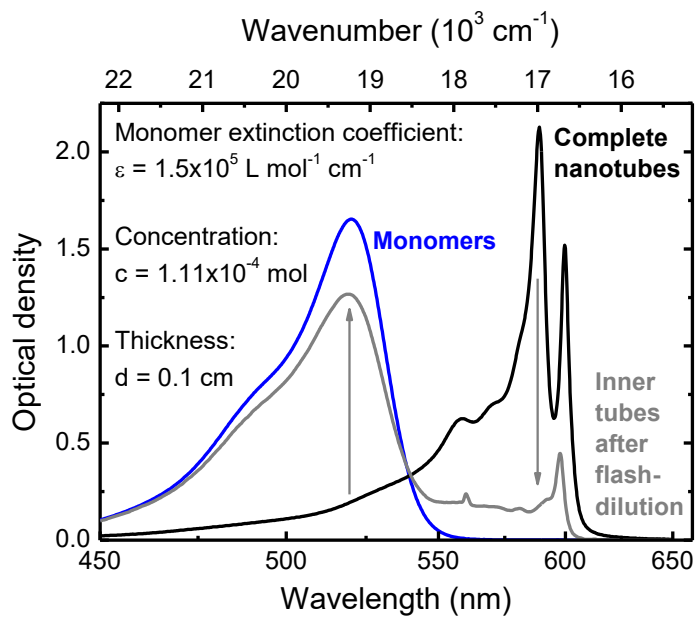

**Supplementary Figure 1.** Absorption spectra of C8S3 monomers (blue), complete nanotubes (black) and flash-diluted inner tubes (gray). The arrows indicate the main spectral changes upon flash-dilution, i.e., dissolution of the outer layer. The peak monomer extinction coefficient is specified by the supplier (FEW chemicals, Wolfen, Germany) as $\epsilon = 1.5 \times 10^5 \ M^{-1} \ cm^{-1}$. In the experiment, the molar concentration of the sample was $c = 1.11 \times 10^{-4} \ M$ and the cuvette thickness $d = 0.1$ cm.

Starting with the former, an upper estimate for the monomer absorption is found by assuming that all molecules ($c = 1.11 \times 10^{-4}$ $M$) are dissolved. In that case the expected optical density amounts to $OD_{max} = \epsilon\, c\, d = 1.66$, where $\epsilon$ is the extinction coefficient of dissolved C8S3 molecules. In the experiment, however, an optical density of only $OD_{exp} = 1.27$ is observed upon flash-dilution. The ratio of these optical densities of $OD_{exp}/OD_{max} = 0.77$ thus indicates that 77% of the molecules were dissolved due to flash-dilution. This, in turn, leaves a concentration of $c = 2.6 \times 10^{-5}$ M for the molecules that still reside in a nanotube after flash-dilution.

The second estimate for the molar concentration is based on the fact that about 60% and 40% of the molecules reside in the outer and inner layer, respectively[2]. Therefore, in case of perfect flash-dilution, where all inner tubes stay intact, one expects a monomer concentration of $c = 6.67 \times 10^{-5}$ $M$, which would lead to an $OD \approx 1$ at 520 nm. This has to be considered as a lower limit of the monomer absorption and clearly underestimates this contribution under the experimental conditions, where the discrepancy arises from the complete dissolution of nanotubes. In fact, the main absorption peak of the inner tubes (~599 nm) decreases by a factor of ~0.7, which indicates that only ~30% of the nanotubes survive flash-dilution. Including this additional rescaling factor, one finds $c = 1.33 \times 10^{-5}$ $M$, which is in good agreement with the estimate from the monomer absorption. For the calculation of the exciton density, we use the average of both concentrations: $c = (1.94 \pm 0.64) \times 10^{-5} M$.

The low-energy main transition of the isolated inner tubes appears blue-shifted by ~50 cm$^{-1}$ relative to the corresponding transition in case of complete nanotubes, which is consistent with earlier findings from bulk flash-dilution experiments reported in literature[2]. It has previously been shown that the nanotubes' absorption spectrum depends critically on the tube radius[3] so that we hypothesize that stripping of the outer layer leads to slight inflation of the inner tubes' radius, which in turn causes the blue-shift.

## Supplementary Note 2: Integration of the Absorptive and EEI Signals

In order to retrieve the absorptive and EEI transients for isolated inner tubes as well as complete nanotubes (Figures 3 and 4 in the main text; Supplementary Figure 17), the 2D spectra were integrated in the rectangular regions of interest as depicted in Figure 2b in the main text. The exact integration intervals are specified in Supplementary Table 1.

**Supplementary Table 1.** Integration intervals for the absorptive and EEI signal transients of isolated inner and complete nanotubes.

| | | **Absorptive signal** | **EEI signal** |
|---|---|---|---|
| Isolated inner tubes | Inner tube diagonal peak | Exc. [16625 cm$^{-1}$, 16875 cm$^{-1}$] <br> Det. [16680 cm$^{-1}$, 16780 cm$^{-1}$] | Exc. [33375 cm$^{-1}$, 33625 cm$^{-1}$] <br> Det. [16680 cm$^{-1}$, 16780 cm$^{-1}$] |
| Complete nanotubes | Outer tube diagonal peak | Exc. [16925 cm$^{-1}$, 17175 cm$^{-1}$] <br> Det. [16900 cm$^{-1}$, 17000 cm$^{-1}$] <br> (or [17000 cm$^{-1}$, 17100 cm$^{-1}$]) | Exc. [33975 cm$^{-1}$, 34225 cm$^{-1}$] <br> Det. [16900 cm$^{-1}$, 17000 cm$^{-1}$] <br> (or [16950 cm$^{-1}$, 17050 cm$^{-1}$]) |
| | Inner tube diagonal peak | Exc. [16585 cm$^{-1}$, 16835 cm$^{-1}$] <br> Det. [16600 cm$^{-1}$, 16700 cm$^{-1}$] <br> (or [16650 cm$^{-1}$, 16750 cm$^{-1}$]) | Exc. [33295 cm$^{-1}$, 33545 cm$^{-1}$] <br> Det. [16600 cm$^{-1}$, 16700 cm$^{-1}$] <br> (or [16650 cm$^{-1}$, 16750 cm$^{-1}$]) |
| | Cross peak (outer → inner) | Exc. [16925 cm$^{-1}$, 17175 cm$^{-1}$] <br> Det. [16600 cm$^{-1}$, 16700 cm$^{-1}$] <br> (or [16650 cm$^{-1}$, 16750 cm$^{-1}$]) | Exc. [33975 cm$^{-1}$, 34225 cm$^{-1}$] <br> Det. [16600 cm$^{-1}$, 16700 cm$^{-1}$] <br> (or [16650 cm$^{-1}$, 16750 cm$^{-1}$]) |

In practice, vertical slices of the 2D spectra were averaged over 250 cm$^{-1}$ (corresponding to three data points) along the excitation axis. Next, the baseline was subtracted from these vertical slices and the respective signal of interest was averaged along the detection axis over 100 cm$^{-1}$ (corresponding to 10 data points). Due to the increased number of features in the absorptive 2D and EEI2D spectra in the case of complete nanotubes, individual contributions from GSB/SE and ESA with opposite signs are more likely to spectrally overlap and, hence, partially compensate each other. At the highest exciton density and, hence, the strongest signals we found this partial compensation to lead to peak shifts, which we accounted for by slightly adjusting the integration area (specified in parenthesis in Supplementary Table 1) in order to avoid simultaneous integration over negative and positive signals.

One of the dominant sources of uncertainty of the extracted signal amplitudes were fluctuations of the background due to unsuppressed scattering of the pump and probe pulses. We determine the standard error of these background fluctuations during each measurement (i.e., at a given exciton density) for the respective spectral regions of interest for the absorptive and EEI signals. The same excitation frequency limits (Supplementary Table 1) are used as before from which the background signal is extracted for each waiting time in the spectral interval from 16000 $cm^{-1}$ to 16200 $cm^{-1}$ along the detection axis. The error bars are identical for all waiting times within the same scan, but may be slightly different for the absorptive and EEI signals.

# Supplementary Note 3: 2D Spectra for Other Exciton Densities and Waiting Times

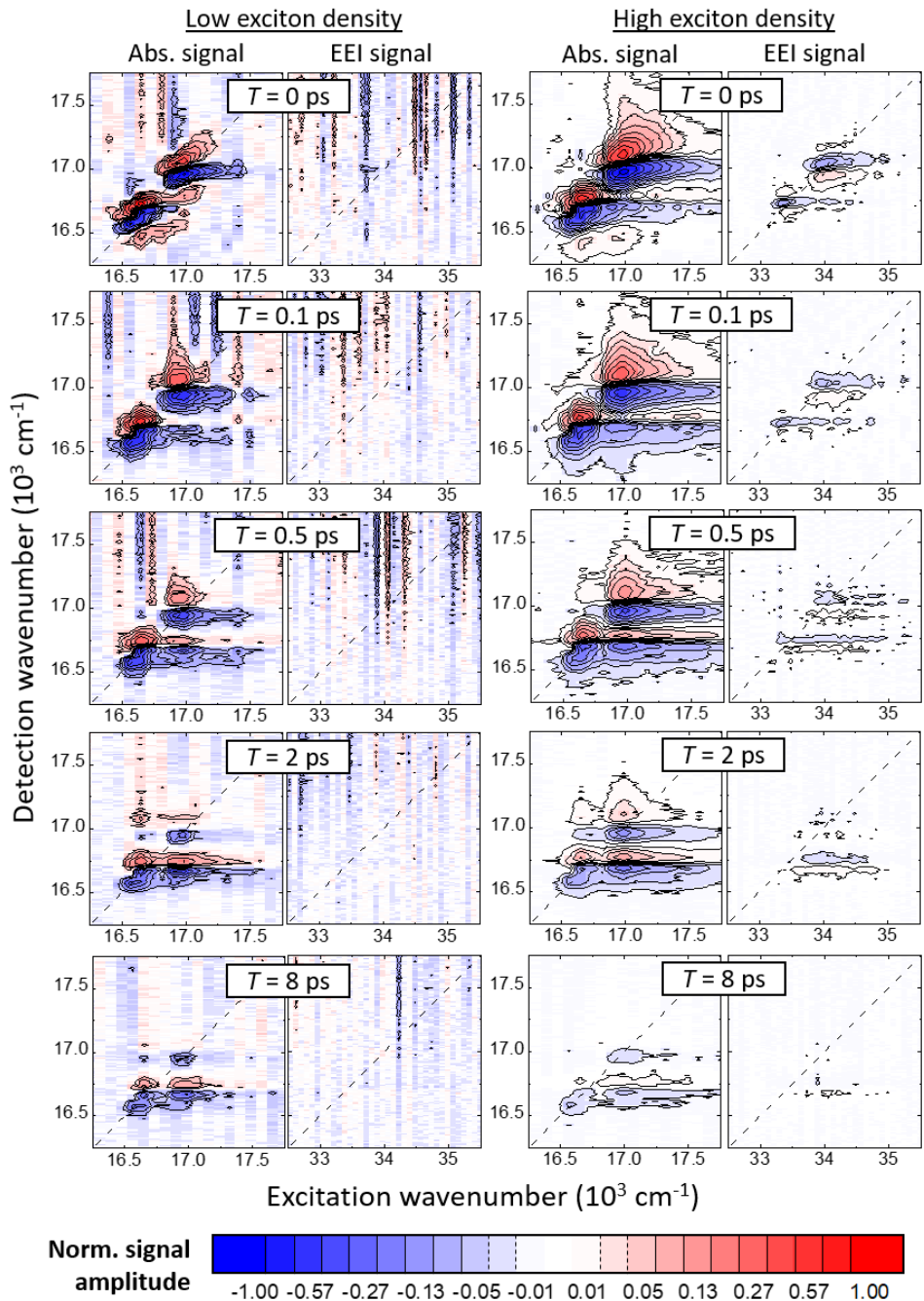

**Supplementary Figure 2.** Absorptive 2D and EEI2D spectra of complete nanotubes recorded at low (1 exciton per ~600 molecules; left column) and high (1 exciton per ~60 molecules; right column) exciton

densities for a range of waiting times. All shown spectra are normalized to the maximum absolute amplitude of the respective absorptive signal at zero waiting time, which preserves the relative scaling between the absorptive and EEI signals. The signal amplitude is depicted on a color scale (between -1 and +1) with increments at 0.83, 0.57, 0.4, 0.27. 0.19, 0.13, 0.08, 0.05, 0.03, and 0.01 to ensure visibility of all peaks at all waiting times. For the spectra at high exciton density all contour lines are drawn, whereas for the low exciton density spectra the contour lines of the lowest levels are omitted as indicated on the color bar (dashed lines are not used for low exciton density).

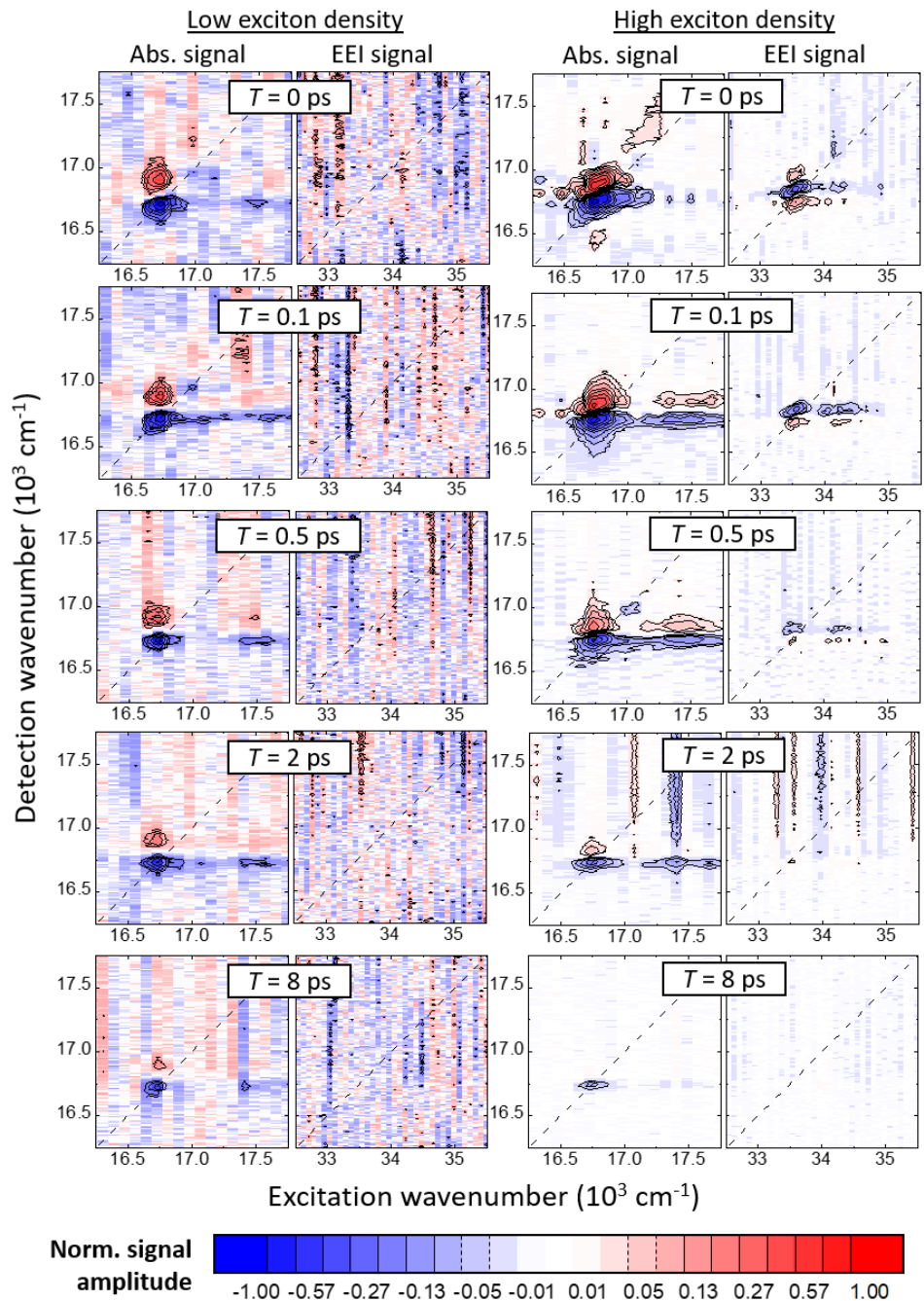

**Supplementary Figure 3.** Absorptive 2D and EEI2D spectra of isolated inner tubes recorded at low (1 exciton per ~400 molecules; left column) and high (1 exciton per ~20 molecules; right column) exciton densities for a range of waiting times. All shown spectra are normalized to the maximum absolute amplitude of the respective absorptive signal at zero waiting time, which preserves the relative scaling between the absorptive and EEI signals. The signal amplitude is depicted on a color scale (between -1 and

+1) with increments at 0.83, 0.57, 0.4, 0.27. 0.19, 0.13, 0.08, 0.05, 0.03, and 0.01 to ensure visibility of all peaks at all waiting times. The drawn contour lines are indicated on the color bar (dashed lines are not used for low exciton density).

## Supplementary Note 4: Absorptive and EEI Cross Peaks from Intra-Band Relaxation

In the case of isolated inner tubes, weak cross peaks can be identified in the absorptive 2D and EEI2D spectra at the detection frequency of the inner tubes ($\omega_{inner}$) at higher excitation frequencies. The appearance of these cross peaks is linked to one of the blue-shifted transitions of the nanotube absorption spectrum (Figure 1a, main text), which originates from the complex molecular packing with two molecules per unit cell[2]. In fact, each molecule in the unit cell gives rise to two excitonic transitions, one of which is polarized parallel and the other orthogonal to the nanotubes' long axis[4]. As a result, the absorption spectrum of the inner tubes comprises a total of four transitions, out of which only the parallel polarized transitions at ~16750 cm$^{-1}$ and ~17500 cm$^{-1}$ are relevant for 2D spectroscopy due to polarization-selective excitation. The latter was facilitated by the polarization of the excitation pulses set parallel to the sample flow along which the nanotubes preferentially align due to their large aspect ratio.

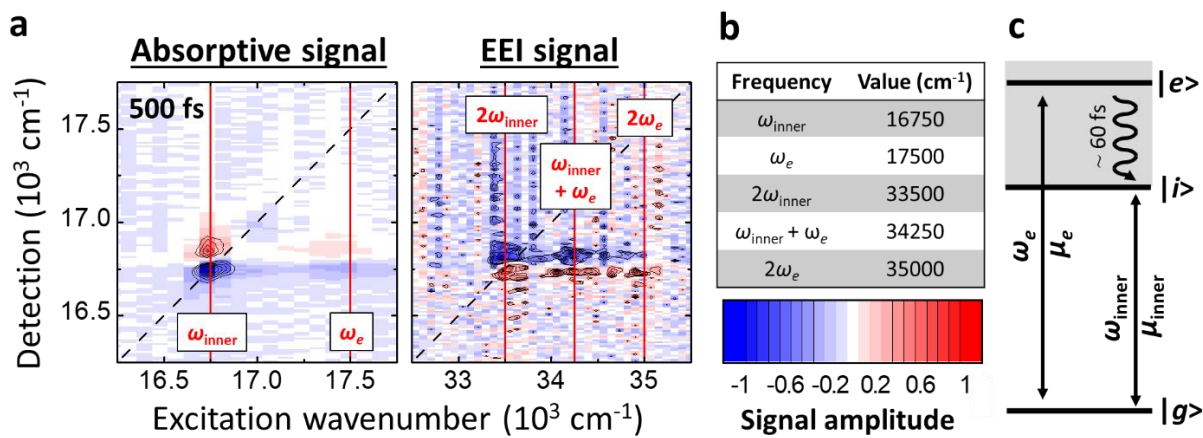

**Supplementary Figure 4.** Absorptive and EEI cross peak for isolated inner tubes. (**a**) Representative absorptive 2D (left) and EEI2D (right) spectra recorded at a waiting time of 500 fs for isolated inner tubes at the highest exciton density of one exciton per ~20 molecules. The spectra were normalized to their maximum absolute amplitude. Dashed lines are drawn at $\omega_{\text{excitation}} = \omega_{\text{detection}}$ and $\omega_{\text{excitation}} = 2\omega_{\text{detection}}$ for absorptive 2D and EEI2D spectra, respectively. (**b**) Summary of the relevant excitation frequencies of optical transitions for isolated inner tubes for absorptive and EEI signals. These frequencies are shown as vertical red lines in the 2D spectra. (**c**) Level diagram of isolated inner tubes with the intra-band exciton state ($|e\rangle$) explicitly drawn. Optical transitions are depicted as vertical arrows with the corresponding frequencies and transition dipole moment indicated. Intra-band exciton relaxation is shown as a wiggly arrow.

The states corresponding to the strong transition at $\omega_{\text{inner}} \sim 16750$ cm$^{-1}$ are situated at the bottom of the exciton band, i.e., the super-radiant states[5], for which an extensive analysis is presented in the main part of the paper. In contrast, the high-frequency transition corresponds to states that lie deep within the exciton band (Supplementary Figure 4c), which we denote as $|e\rangle$ with the corresponding frequency $\omega_e$ and transition dipole moment $\mu_e$. Excitation of this transition is followed by ultrafast intra-band relaxation on a sub-100 fs timescale[6], which leads to additional population of the bottom states of the exciton band encoded in a rapidly in-growing cross peak in the absorptive 2D spectra ($\omega_e \rightarrow \omega_{inner}$). Note that in our experiments the corresponding diagonal peak could hardly be detected because of its short-lived nature and sparse sampling of the waiting time. However, previously published transient absorption (TA) data revealed a decay time as short as ~60 fs for this transition[7]. An additional complication in measuring the diagonal peak arises from the fact that its amplitude scales with the already small dipole moment ($|\mu_e|^4$), whereas the cross peak involves the stronger transition dipole moment of the inner tube ($|\mu_e|^2|\mu_{inner}|^2$) and is therefore easier to detect. In comparison, the same cross peak ($\omega_e \rightarrow \omega_{inner}$) is present in the absorptive 2D spectra of complete nanotubes, but only visible as peak elongations towards higher excitation frequencies (Figure 2b in the main text), as they partially overlap with the much stronger cross peak due to the outer layer.

Following ultrafast intra-band relaxation, excitons can further diffuse and eventually undergo exciton–exciton annihilation, which is reflected in the emergence of cross peaks in the EEI2D spectra (Supplementary Figure 4a). Specifically, the strongest cross peak is observed at a detection frequency $\omega_{inner}$ and an excitation frequency of 34250 cm$^{-1}$ (marked by the center vertical line at $\omega_{inner} + \omega_e$ in Supplementary Figure 4a), which corresponds to the sum of the contributing excitation frequencies, i.e., (16750 + 17500) cm$^{-1}$ (see table in Supplementary Figure 4b). In comparison, the other vertical lines refer to the fundamental transitions at excitation frequencies of $2\omega_{inner}$ and $2\omega_e$. The EEI cross peak ($\omega_{inner} + \omega_e \rightarrow \omega_{inner}$) encodes the mutual interaction of excitons one of which was directly excited at the bottom of the exciton band, whereas the other one underwent intra-band relaxation. Note that for complete nanotubes the spectral region around the detection frequency $\omega_{inner}$ is dominated by the EEI cross peak at double the excitation frequency of the outer tube ($2\omega_{outer} = 34000$ cm$^{-1}$).

# Supplementary Note 5: Double-Sided Feynman Diagrams for the EEI Signals

## 5.1. EEI Diagonal Peaks

In this section we present the double-sided Feynman diagrams that contribute to the EEI signal of the diagonal peaks of the isolated inner tubes (Supplementary Figure 5) and the diagonal peaks of the outer tube (Supplementary Figure 6). For conciseness, only the rephasing diagrams as determined by their phase matching condition, i.e., $k_{\text{signal}} = -2k_{pu} + 2k_{pu} + k_{pr}$, are shown[8], where $k_{pu}$ is the wavevector of the pump beam, $k_{pr}$ the wavevector of the probe beam, and detection occurs along direction $k_{\text{signal}}$. The non-rephasing diagrams can be derived by considering diagrams that emit a signal field in the phase-matched direction of $+2k_{pu} - 2k_{pu} + k_{pr}$. However, a rigorous mathematical analysis, where all signals are computed in the framework of the response function theory and subsequently convoluted with the electric fields of the involved laser pulses is beyond the scope of this work[8,9].

For the EEI diagonal peak of isolated inner tubes we consider diagrams that give rise to a signal at an excitation frequency of $2\omega_{\text{inner}}$ and detection at $\omega_{\text{inner}}$. In the diagrams for the isolated inner tubes $|g\rangle$ represents the electronic ground-state, $|i\rangle$ and $|j\rangle$ the one-exciton states of the inner tube, $|ii\rangle$ and $|jj\rangle$ bi-exciton states, and analogously $|jjj\rangle$ for the tri-exciton states (see the level diagram in Supplementary Figure 5. Note that we formally distinguish between the states $|i\rangle$ and $|j\rangle$ (and $|ii\rangle$ and $|jj\rangle$) of the two neighboring excitons to include the fact that the exciton state after the waiting time $T$ is not necessarily identical to the exciton state prepared by the pump pulses. All diagrams in Supplementary Figure 5 share the general structure that the first four interactions with the pump pulses ($\mp 2k_{pu}$) excite the inner tube followed by the probe pulse ($+k_{pr}$) interacting with the same tube. During the waiting time *T*, exciton–exciton annihilation (EEA) can occur (Supplementary Figure 5, right column; highlighted in orange), if the system is in a bi-exciton population state. As shown in literature the EEI signal is then dominated by EEA[1,10].

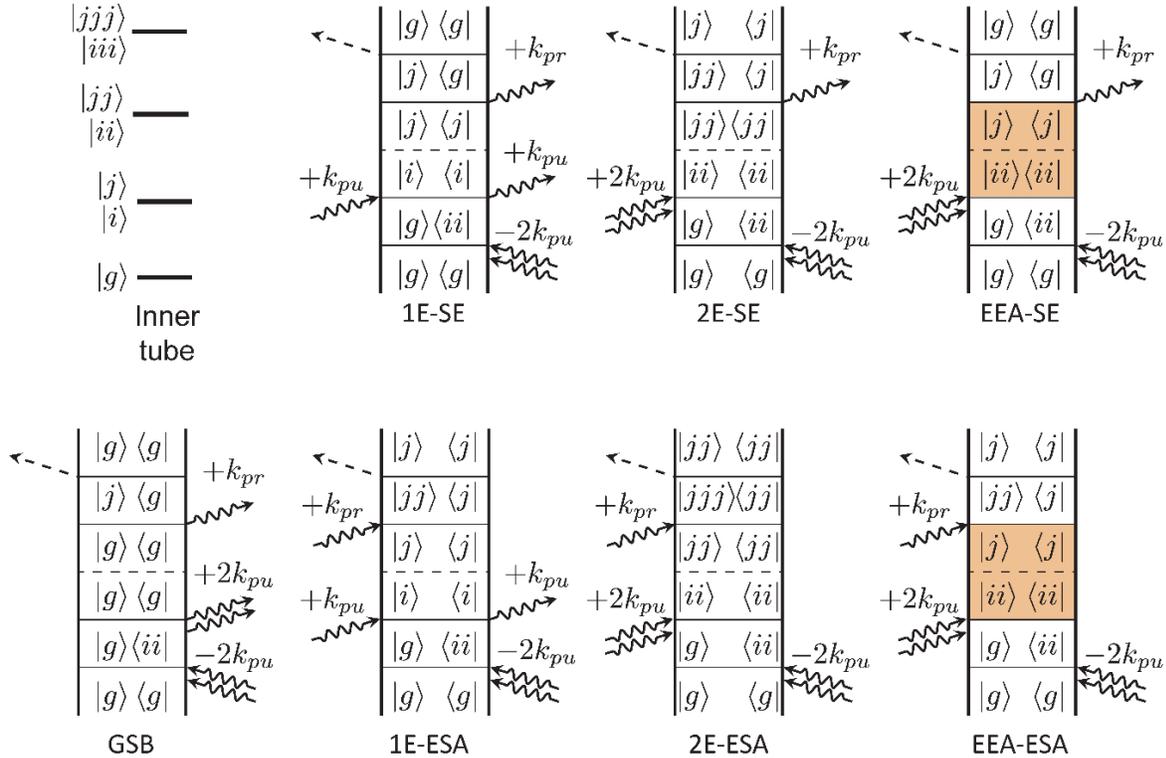

**Supplementary Figure 5.** Rephasing double-sided Feynman diagrams, which contribute to the EEI diagonal peak of isolated inner tubes ($2\omega_{\text{inner}} \to \omega_{\text{inner}}$). The level diagram for the isolated inner tubes is shown in the upper left corner. In the diagrams time flows from the bottom to the top during which the interactions with the laser pulses are indicated by arrows. The dashed line indicates propagation during the waiting time *T*. The double interaction with each of the two pump pulses can create a population of the ground state, a one-exciton (1E) state or a bi-exciton (2E) state, which are subsequently probed by GSB ($|g\rangle \to |j\rangle$), SE ($|j\rangle \to |g\rangle$ or $|jj\rangle \to |j\rangle$) or ESA ($|j\rangle \to |jj\rangle$ or $|jj\rangle \to |jjj\rangle$). The process of exciton–exciton annihilation (EEA) is shaded in orange.

In absence of exciton transfer (ET) between the tubes, the description of the outer tube diagonal peak is identical to the isolated inner tube with exception of the notation of the states. Hence, the former can be obtained by renaming the states according to $|i\rangle \to |o\rangle$, $|ii\rangle \to |oo\rangle$, *etc*. Nevertheless, we explicitly include these diagrams in Supplementary Figure 6 here, as they form the basis for the discussion on the EEI cross peak in the next section.

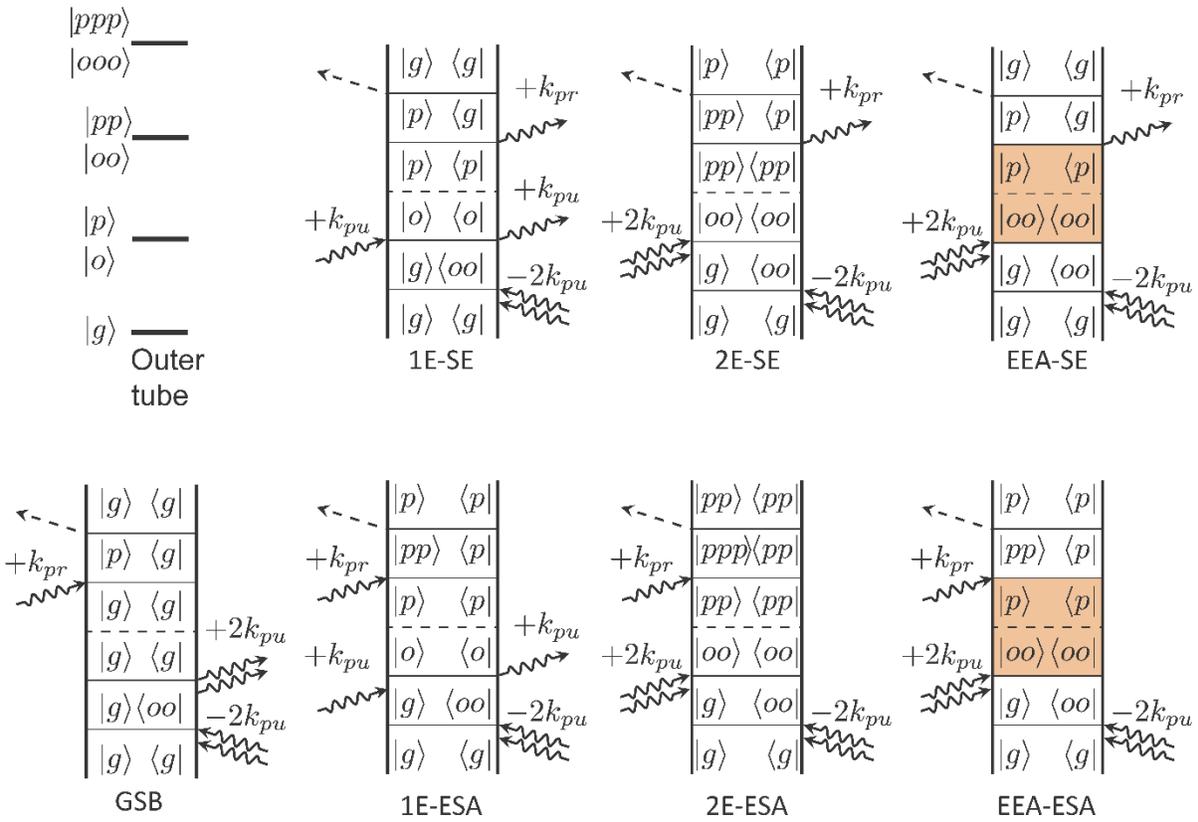

**Supplementary Figure 6.** Rephasing double-sided Feynman diagrams, which contribute to the EEI diagonal peak of the outer tube. The level diagram for the outer tube assuming that the absence of any inter-tube ET, is shown in the upper left corner.

## 5.2. EEI Cross Peak

In order to understand the character of the EEI cross peak we discuss the double-sided Feynman diagrams, which contribute to the particular signal at double the excitation frequency of the outer tube with subsequent detection at the frequency of the inner tube, i.e., $2\omega_{outer} \rightarrow \omega_{inner}$. At time zero ($T = 0$), where neither exciton transfer (ET) nor EEA occur, only three diagrams contribute to the EEI cross peak signal (Supplementary Figure 7, black box). Here, the first four interactions with the pump pulses excite the outer tube, while the probe pulse interacts with the inner tube. Since the outer and inner tubes can be considered as weakly coupled[2], excitation of the outer tube does not influence the inner tube and *vice versa*. Due to the weak coupling we also exclude the possibility of any inter-tube exciton–exciton annihilation, where two excitons residing on different tubes annihilate directly without any ET event involved. Therefore, the state labelled as $|pi\rangle$ (see level diagram in Supplementary Figure 7) refers to the situation of two independent excitons – one on each tube. Analogously, $|ppi\rangle$ describes the situation of one exciton located on the inner and two excitons located on the outer tube. Due to the different overall signs of the diagrams (Supplementary Figure 7, black box) all different pathways mutually compensate each other and, hence, no EEI cross peak is visible. Analogously, it can be shown that for weakly coupled systems the Feynman diagrams corresponding to the absorptive cross peak ($\omega_{outer} \rightarrow \omega_{inner}$) cancel each other at zero waiting time[11].

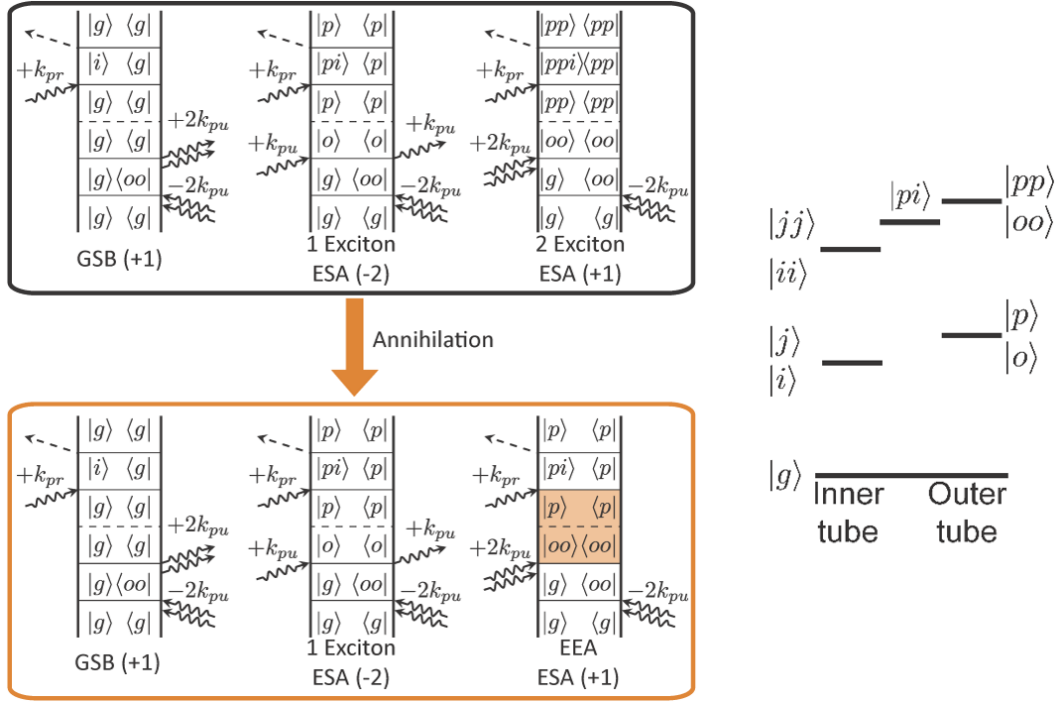

**Supplementary Figure 7.** Rephasing double-sided Feynman diagrams for the EEI cross peak at zero waiting time. Top panel (black): Pathways without inter-tube exciton transfer (ET) and EEA. Bottom panel (orange): Possible pathways including EEA, but still no ET. The level diagram for complete nanotubes is shown on the right side.

EEA on the outer tube in absence of any ET leads to a modified set of diagrams for the EEI cross peak, where the third diagram is replaced by one that contains EEA (Supplementary Figure 7, orange box; diagram on the right). However, under the premise of no ET and assuming weak coupling, EEA on the outer tube does not influence the exciton dynamics on the inner tube and, thus, does not alter the interaction of the probe pulse with the latter. Therefore, the diagrams at zero waiting time still mutually cancel, as it is the case in upper panel of Supplementary Figure 7 and, thus, no EEI cross peak emerges.

For finite waiting times *T,* inter-tube ET has to be considered explicitly. As a result, the condition that excitons on the outer tube will not influence exciton processes on the inner tube does not hold any longer. The corresponding double-sided Feynman diagrams, which contain both contributions, i.e., ET (shaded in blue) and EEA (shaded in orange), are shown in Supplementary Figure 8. As shown in literature the EEI signal is dominated by pathways that include EEA[1,10], although these diagrams co-exist with a number of diagrams that contain only ET, which formally also give rise to a signal at $2\omega_{outer} \rightarrow \omega_{inner}$. The fact that the EEI cross peak dynamics in experiment (see Figure 4 in the main text) exhibit a dependence on the excitation intensity corroborates the fact that the diagrams containing both ET and EEA are most relevant to the EEI cross peak, as for the diagrams with ET alone no intensity dependence is expected.

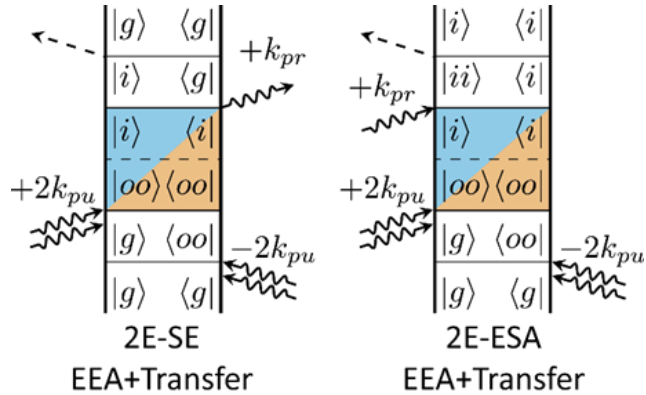

**Supplementary Figure 8.** Rephasing double-sided Feynman diagrams which contribute to the cross peak of the EEI signal. Additional to EEA (orange) population transfer is included (blue).

### 5.3. Seventh-Order Signals

As discussed in the main text, higher-order, namely at least seventh order, effects occur at high exciton densities, which were indeed observed experimentally at three times of the fundamental frequency (Supplementary Note 7). Although the seventh-order signal can be glimpsed spectroscopically isolated at triple the fundamental frequency, it also contributes to the (fifth-order) EEI signal at twice the fundamental frequency. This occurs in a similar fashion as the (fifth-order) EEI signal contributes to the (third-order) absorptive signal, as the transient exciton dynamics accelerate in presence of exciton–exciton annihilation. Two diagrams that demonstrate the effect of the seventh-order on the EEI signal of the diagonal peak of the outer tube (left diagram; $2\omega_{outer} \rightarrow \omega_{outer}$) and the cross peak (right diagram; $2\omega_{outer} \rightarrow \omega_{inner}$) are shown in Supplementary Figure 9a. Two exemplary diagrams for the diagonal peak and the cross peak at three times the excitation frequency $3\omega_{outer}$ are shown in Supplementary Figure 9b.

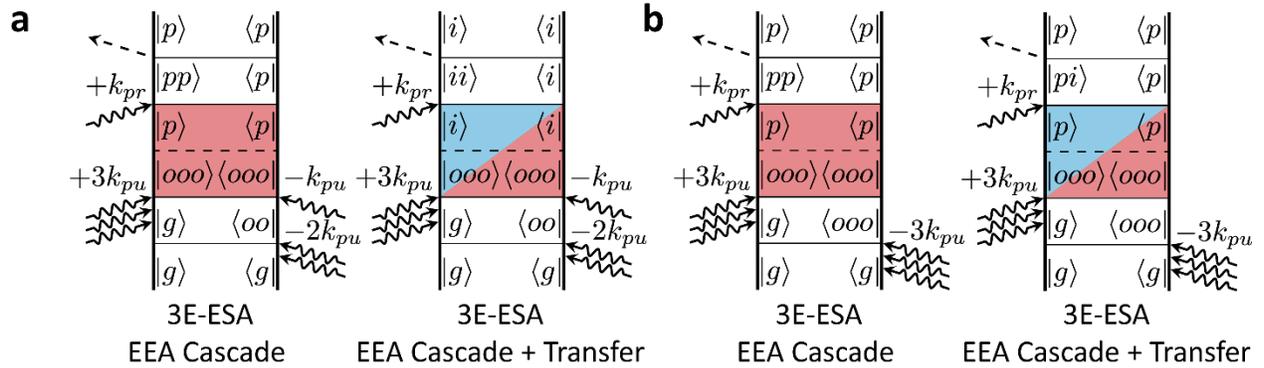

**Supplementary Figure 9.** (**a**) Rephasing double-sided Feynman diagrams of the seventh-order signal which give rise to a signal at an excitation frequency of $2\omega_{outer}$ and detection at $\omega_{outer}$ or $\omega_{inner}$ following ET. EEA cascades are shown in dark red, while ET from the outer tube to the inner tube is shaded in blue. (**b**) Rephasing double-sided Feynman diagrams for the signal at $3\omega_{outer}$.

Possible contributions to the seventh-order signal include a sequential cascade of annihilations, in which excitons participate in multiple EEA events during the waiting time $T$. For example, after a bi-exciton state relaxes to a one-exciton state, it is excited one more time to the bi-exciton state that subsequently relaxes. However, such processes are not very likely in our experiments because of the short (~15 fs) pulse duration, the low intensity of the probe pulse and a single-pump-beam geometry, which prevents the second excitation to occur with a photon from the same pump pulse.

## Supplementary Note 6: Monte-Carlo Simulations

### 6.1. Simulation Grid

The molecular grid for Monte-Carlo (MC) simulations was set up to match the size known from cryo-TEM measurements[2,12,13] and previously published theoretical models[2]. The boundary conditions for the grid are given by the radii and the molecular surface densities of both tubes. For simplicity we assume identical square grids for the inner and outer tube with a single molecule on each grid site, although more sophisticated models for the molecular packing have been proposed including brickwork models[14–16] and extended herringbone models[2,3]. Here, we use the same model parameters as in Ref. [2]: $R_{outer} = 6.465$ nm and $R_{inner} = 3.551$ nm for the radii, $N_{outer} = 14260$ and $N_{inner} = 7992$ for the number of molecules in each tube and a total tube length of $L \approx 197$ nm. From these values the molecular surface densities are calculated as $\rho = N/2\pi R L$. In order to construct the square grid for MC simulations we use the average molecular surface density of both layers ~1.81 molecules nm$^{-2}$, from which the lattice constant is calculated as $a = \sqrt{1/\rho} \approx 0.74$ nm. The number of molecules on the circumference then follows as $N_c = 2\pi R/a$. A summary of the relevant parameters for the molecular grid is given in Supplementary Table 2.

**Supplementary Table 2.** Summary of parameters for the molecular grid for MC simulations. The chemical structure of C8S3 is shown in the right column with dimensions indicated. Note that the latter refers to the situation with the octyl sidegroups fully extended.

| Quantity | Inner layer | Outer layer | Chemical structure |
|---|---|---|---|
| Radius (Ref. [2]) | 3.551 nm | 6.465 nm | |
| Circumference | ~22.3 nm | ~40.6 nm | |
| Molecular surface density | ~1.84 molecules nm$^{-2}$ | ~1.77 molecules nm$^{-2}$ | |
| Lattice constant | ~0.74 nm | | |
| Unit cell area | ~0.55 nm$^2$ | | |
| Number of molecules on the circumference | ~30 | ~55 | |

   At first glance the value for the lattice constant disagrees with the molecular geometry, as the size of the molecule exceeds this length (see Supplementary Table 2 for chemical structure). However, it is important to realize that assuming a simple square grid for the molecular packing yields a single effective lattice constant, which averages the actual separation between individual molecules in different directions for more sophisticated packing motifs. In reality, the molecules are expected to stack in one direction with their chromophores closely aligned at distances on the order of 0.4 nm (as reported for a structurally similar molecule[17]), while the molecular separation in the "lateral" direction would roughly correspond to the molecular size of 1.7 nm. Combined this yields a unit cell area of 1.7 nm × 0.4 nm = 0.68 nm$^2$, which agrees well with the unit cell area derived from the square grid. The remaining molecular dimension (~1.7 nm) sticks out perpendicular from the planes considered here and contributes to the wall thickness of the double-walled assembly. According to the inner and outer radii the thickness amounts to ~3 nm, which is in good agreement with two times the molecular size in this direction.

A schematic representation of the molecular grid used for MC simulations comprising two planes for the inner (red) and outer (gray) tube is shown in Supplementary Figure 10. For simulations of the exciton dynamics of the isolated inner tubes, only the bottom plane was used. Excitons are depicted as orange circles in order to visualize their annihilation radius. In the MC simulations excitons can perform the following processes: (1) decay according to their lifetime, (2) hop between adjacent sites, (3) vertically transfer between the two layers and (4) undergo exciton–exciton annihilation. The latter occurred, when two excitons were mutually overlapping within their annihilation radius, as exemplarily shown on the outer layer.

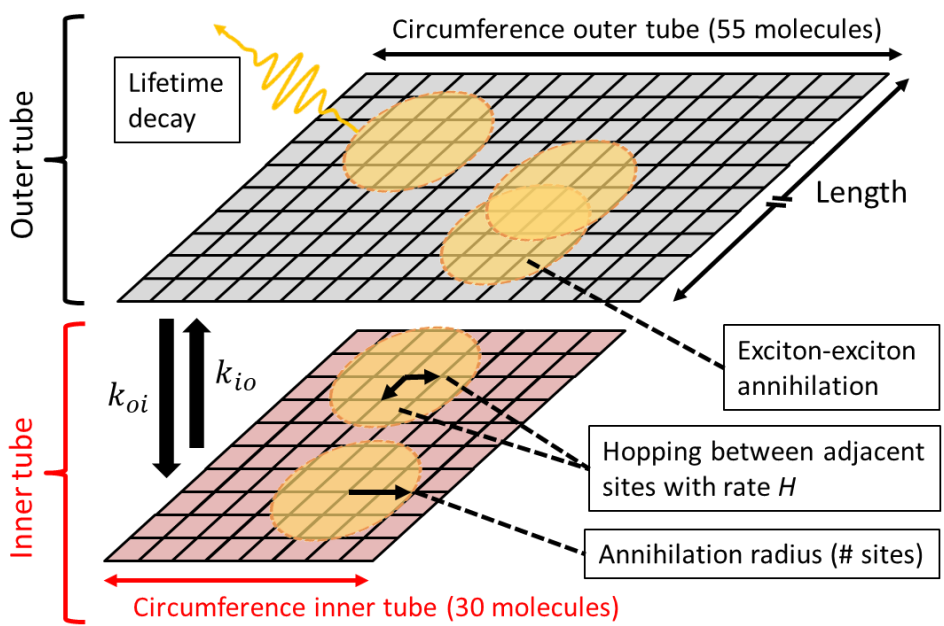

**Supplementary Figure 10.** Molecular grid for Monte-Carlo simulations. The inner and outer tube are depicted as planes shaded in red and gray, respectively. Excitons are shown as orange circles with their size corresponding to the annihilation radius. The different processes that excitons can undergo during the MC simulations are exemplarily shown.

## 6.2. Extraction of the Absorptive and EEI Signal

**Supplementary Table 3.** Prerequisites for extraction of the absorptive and EEI signals from MC simulations.

| Signal | Peak type | Origin | Annihilated? | Position at time t | Freq. (exc.→det.) |
|---|---|---|---|---|---|
| Absorptive | diagonal peak | inner | - | inner | $\omega_{inner} \rightarrow \omega_{inner}$ |
| Absorptive | diagonal peak | outer | - | outer | $\omega_{outer} \rightarrow \omega_{outer}$ |
| Absorptive | cross peak | outer | - | inner | $\omega_{outer} \rightarrow \omega_{inner}$ |
| EEI | diagonal peak | inner inner | yes | inner | $2\omega_{inner} \rightarrow \omega_{inner}$ |
| EEI | diagonal peak | outer outer | yes | outer | $2\omega_{outer} \rightarrow \omega_{outer}$ |
| EEI | cross peak | outer outer | yes | inner | $2\omega_{outer} \rightarrow \omega_{inner}$ |

## 6.3. Definition of the Annihilation Radius

We use the annihilation radius as a quantity to characterize the distance dependence of the interactions of two approaching excitons that ultimately results in annihilation of one of the excitons. If we assume a Förster-type exchange for an exciton–exciton annihilation event[18–20], the probability of the event scales with distance $R$ as $(1 + R^6/R_0^6)^{-1}$, where $R_0$ is the Förster radius. In order to ease the computations, we approximate this dependence by a step function: the probability of annihilation within $R_0$ is unity, otherwise it is null (Supplementary Figure 11). In other words, within the cut-off distance $R_0$ the annihilation event outcompetes any other relevant rate in the system (e.g. the exciton decay rate and the exciton diffusion rate).

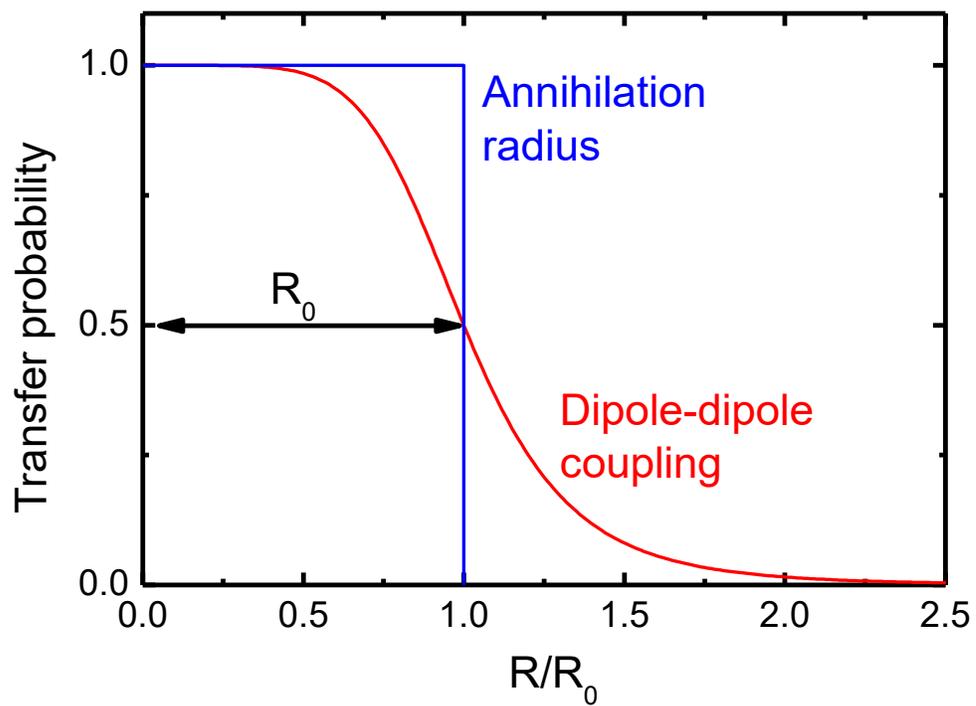

**Supplementary Figure 11.** Distance dependence of the Förster-type transfer probability (red) and its approximation with a step function (blue; here denoted as $R_0$).

## 6.4. Overview of Parameters

**Supplementary Table 4.** Overview of parameters for Monte-Carlo simulations of the exciton dynamics for isolated inner tubes and complete nanotubes.

| Quantity | Symbol | Inner tubes | Complete nanotubes | Source |
|---|---|---|---|---|
| One-exciton lifetime | $\tau$ | 58 ps | 33 ps | PL measurements; Supplementary Note 10 |
| Annihilation radius | $R_0^{\text{inner}}$ $R_0^{\text{outer}}$ | 3 molecules - | 3 molecules 3 molecules | Global fitting parameter; Supplementary Note 6.3 |
| Initial exciton density (number of molecules per exciton) | $\frac{N_m}{N_e}$ | 26 57 170 580 | 14 87 853 | Obtained from excitation flux; varied within uncertainty (Methods and Supplementary Note 1 |
| Molecular grid size | Inner Outer | $30 \times 1000$ - | $30 \times 1000$ $55 \times 1000$ | Derived from model in Ref. [2]; Supplementary Note 6.1 |
| Lattice constant | $a$ | 0.74 nm | 0.74 nm | Derived from model in Ref. [2]; Supplementary Note 6.1 |
| Exciton transfer rate (inner → outer) (outer → inner) | $k_{\text{io}}$ $k_{\text{oi}}$ | - - | 0.0013 fs$^{-1}$ 0.0031 fs$^{-1}$ | Obtained from 2D experiments; Supplementary Note 15 |
| Hopping rate | $H_{\text{inner}}$ $H_{\text{outer}}$ | 0.04 fs$^{-1}$ - | 0.04 fs$^{-1}$ 0.04 fs$^{-1}$ | Global fitting parameter |
| Diffusion constant | $D_{2D}$ | 10 mol. ps$^{-1}$ 5.5 nm$^2$ ps$^{-1}$ | 10 mol. ps$^{-1}$ 5.5 nm$^2$ ps$^{-1}$ | Exciton mean square displacement; Supplementary Note 6.5 |

In the MC simulations only the exciton hopping rate (i.e., the probability of of an exciton to move to any of the neighboring molecules during one timestep in the simulation) and the annihilation radius were treated as free parameters, while all other parameters were fixed as their values were obtained from supplementary experiments or calculations. The exciton density was taken from the experimental conditions and allowed to vary within the experimental uncertainty. The lifetime of a single exciton was measured in time-resolved photoluminescence (PL) experiments under extremely low exciton densities of less than 1 exciton per $10^4$ molecules (Supplementary Note 10). The transfer rate from the outer to the inner tube was measured using conventional 2D spectroscopy (Supplementary Note 15) and agrees with the values from literature[7,21,22]. The opposite rate (inner → outer) follows from the condition that the inner and outer tube exciton populations eventually reach thermal equilibrium, where the net inter-tube transfer rates are identical[23,24]. Hence, this rate is scaled with the Boltzmann factor ($\exp\left(-\frac{\Delta E}{k_B T}\right) \approx 0.22$; with $\Delta E = 300$ cm$^{-1}$ as the energy difference between inner and outer tube and $k_b T \approx 200$ cm$^{-1}$ at room temperature) and the density-of-states. The latter is proportional to the number of molecules in the inner and outer layer, which scales with the tube radii assuming identical molecular surface densities (Supplementary Note 6.1). Taken together one finds a ratio of ~0.4 between the upward and the downward exciton transfer rates.

### 6.5. Exciton Displacement

In order to compute the mean (square) exciton displacement Monte-Carlo simulations were conducted in an annihilation-free setting, i.e., with EEA switched off, and at a low exciton density. The latter was important to not hinder the exciton motion by having too many occupied sites. All excitons were labelled with their initial [$X_i$, $Y_i$, $Z_i$] and final [$X_f$, $Y_f$, $Z_f$] position on the grid, where the *X* and *Y* coordinates refer to sites along and across the molecular grid, respectively. The *Z* coordinate encodes whether an exciton resides on the inner or outer layer for which *Z* can take values of 0 or 1. As inter-layer exciton transfer is constrained to occur vertically, i.e., between corresponding sites on the inner and outer layer, the *Z*-coordinate can be neglected for the calculation of the (square) displacement. Instead, the displacement $x_n$ and square displacement $x_n^2$ of an individual exciton are computed via:

$$x_n = a \sqrt{[(X_f - X_i)^2 + (Y_f - Y_i)^2]},$$

$$x_n^2 = a^2[(X_f - X_i)^2 + (Y_f - Y_i)^2].$$

Here, $a$ is the lattice constant (Supplementary Note 6.1). The histograms for the displacement and square displacements for isolated inner tubes and complete nanotubes are shown in Supplementary Figure 12. In the simulations of the isolated inner tubes and complete nanotubes we find mean square exciton displacements ($<x^2>$) of 1282 nm$^2$ and 722 nm$^2$, respectively, which translate into 2319 molecules and 1307 molecules assuming a molecular surface density of 1.81 molecules nm$^{-2}$. These values differ despite identical exciton hopping rates due to different one-exciton lifetimes, which permits excitons to diffuse longer (and farther) in the case of isolated inner tubes. From the mean square displacement the diffusion constants are calculated ($<x^2> = 4D_{2D}\tau$)$^{25}$. Note that this diffusion constant refers to the situation of isotropic exciton transport based on the underlying (simplified) molecular square grid, where identical exciton hopping rates in all directions are assumed.

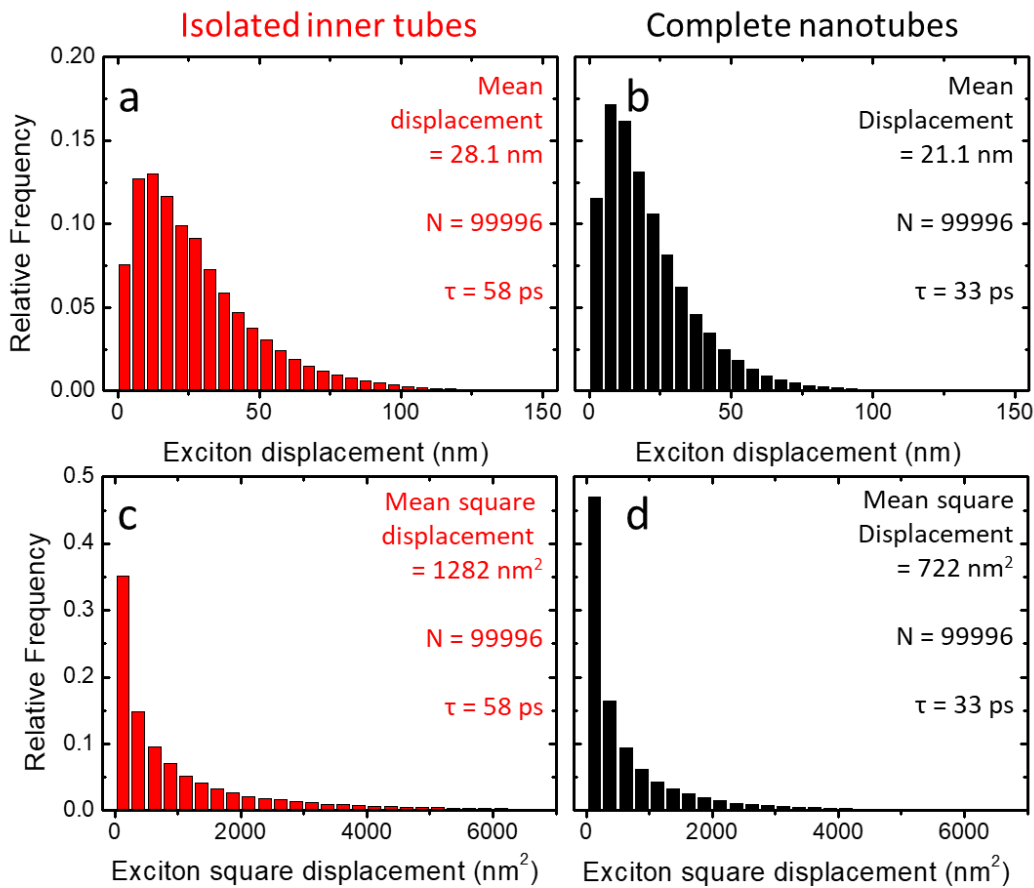

**Supplementary Figure 12.** Exciton (square) displacement in Monte-Carlo simulations. (**a**) and (**b**) histograms for the exciton displacement, and (**c**) and (**d**) square exciton displacement obtained from Monte-Carlo simulations in case of isolated inner tubes (left, red) and complete nanotubes (right, black). For the histograms, the binning width was set to 5 nm in the case of exciton displacement, and 250 nm$^2$ for the exciton square displacement. The inset states the mean (square) exciton displacement, number of excitons the statistics are based on and the one-exciton lifetime.

### 6.6. Exciton–Exciton Annihilation Statistics: Isolated Inner Tubes

In the Monte-Carlo simulations the path of each individual exciton is recorded, during which it could either naturally decay due to its finite lifetime or undergo EEA with another exciton. In that case, one of the excitons is deleted, while the surviving exciton can continue to diffuse and engage in additional EEA events. For the latter, the accumulated number of participations in EEA events was recorded until and including an exciton's own relaxation. As the excitons are not constrained from participation in multiple EEA events, this value may exceed one.

The nature of EEA in which one exciton is destroyed imposes a lower limit on the share of excitons for each number of EEA events. For example, in the extreme case of complete annihilation of all excitons, 50% of the excitons can accumulate a maximum of one annihilation event, 25% of the excitons a maximum of two annihilation events, *etc.,* as described by the geometric distribution. In contrast, in complete absence of exciton–exciton annihilation, excitons decay only according to their lifetime and, thus, accumulate no annihilation events. These two limiting cases dictate the lower and upper bound for the mean number of EEA participations ($<N_{ann}>$) between 0 for the annihilation free case and 2 for complete annihilation of all excitons. Note that these boundaries only hold for a closed system, i.e., the system does not receive any additional excitons from an external source, as for example via exciton transfer from the outer tube.

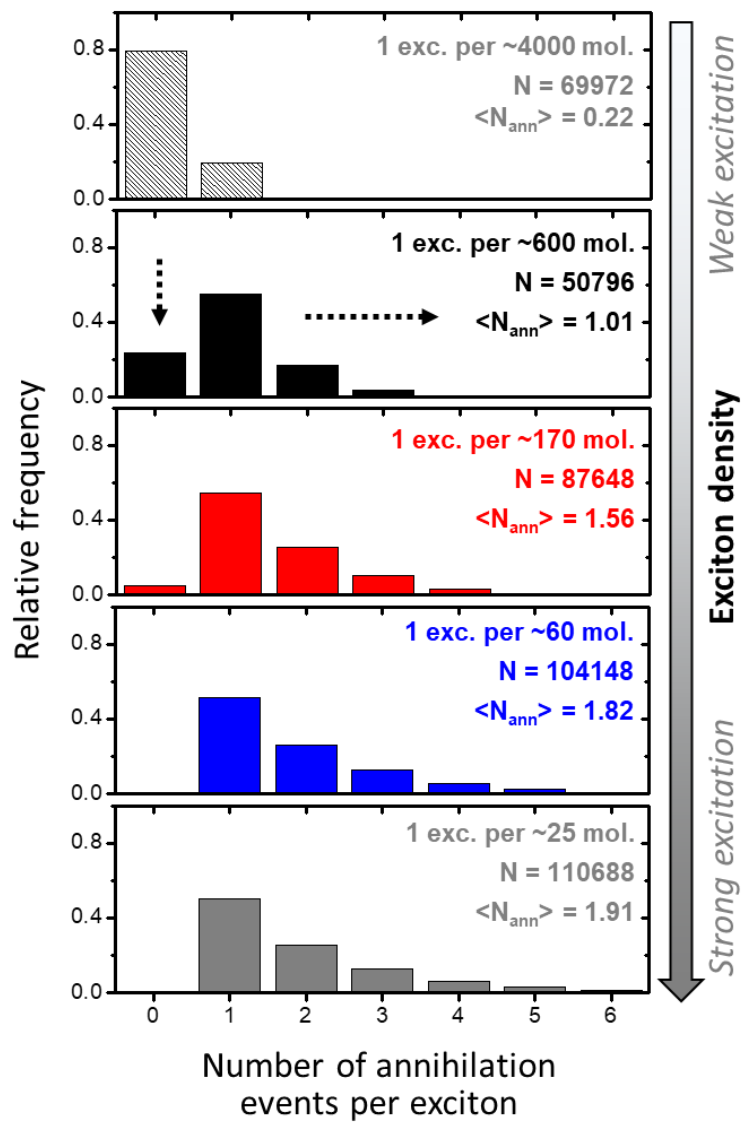

**Supplementary Figure 13.** Histograms for the number of accumulated annihilation events per exciton in the Monte-Carlo simulations of isolated inner tubes at different exciton densities (shown in the inset of each panel). The inset also specifies the number of excitons the statistics are based on and the mean number of annihilation events for the given histogram. The dashed arrows indicate the main changes upon increasing the exciton density. The upper panel does not refer to any exciton density in experiment, but was added to illustrate 20% probability of EEA even at exciton densities several times lower than experimentally used.

Supplementary Figure 13 depicts histograms for the number of annihilation events that excitons accumulated during the simulation of isolated inner tubes for a range of exciton densities. Increasing the exciton density leads to more prominent EEA. As a result, there is a lower number of excitons that decay naturally, which is reflected in a decreasing number of excitons that did not participate in any annihilation event, i.e., the number of accumulated EEA events of zero. Simultaneously, the distribution shifts to higher numbers of EEA events, as excitons are more likely to encounter another exciton and, thus, engage in another EEA event. Expectedly, the mean number of EEA participations increases from 1.01 at one exciton per ~600 molecules up to 1.91 at one exciton per ~20 molecules for increasing exciton densities evidencing the importance of multi-annihilation events accumulated by individual excitons. Even for the lowest exciton density in experiment, MC simulations show that a considerable share of the excitons has accumulated two EEA events by the time of their death, which is the primary requirement for the observation of multi-exciton processes encoded in seventh and higher-order signals.

### 6.7. Exciton–Exciton Annihilation Statistics: Complete Nanotubes

In order to elucidate the fate of the excitons that were originally planted on the outer tube, we extract the fraction of these excitons that (1) decay naturally, (2) decayed due to EEA on the inner tube or (3) decayed due to EEA on the outer tube at various exciton densities as shown in Supplementary Figure 14. At low exciton densities, the inter-layer exciton transfer (ET) rate is significantly faster than the EEA rates (i.e., number of EEA events per time interval), which causes the majority of the excitons from the outer tube to be transferred first and then decay naturally or, to a lesser extent, annihilate on the inner tube (Supplementary Figure 14, red curve).

At the same time, EEA on the outer tube is negligibly small (Supplementary Figure 14, black curve). In contrast, for very high exciton densities, the EEA rate accelerates and ultimately outcompetes the ET rate, which leads to very prominent EEA on the outer tube. As a result, the exciton population is depleted before any of the surviving excitons can be transferred to the inner tube for which EEA is then less pronounced. At intermediate exciton densities, the ET rate and EEA rate are balanced, which maximizes the share of excitons that were planted on the outer tube, but annihilate on the inner tube. Note that Figure 5 in the main text is a sum of the excitons that either decayed naturally (Supplementary Figure 14; green squares) or annihilated (Supplementary Figure 14; red triangles) on the inner tube.

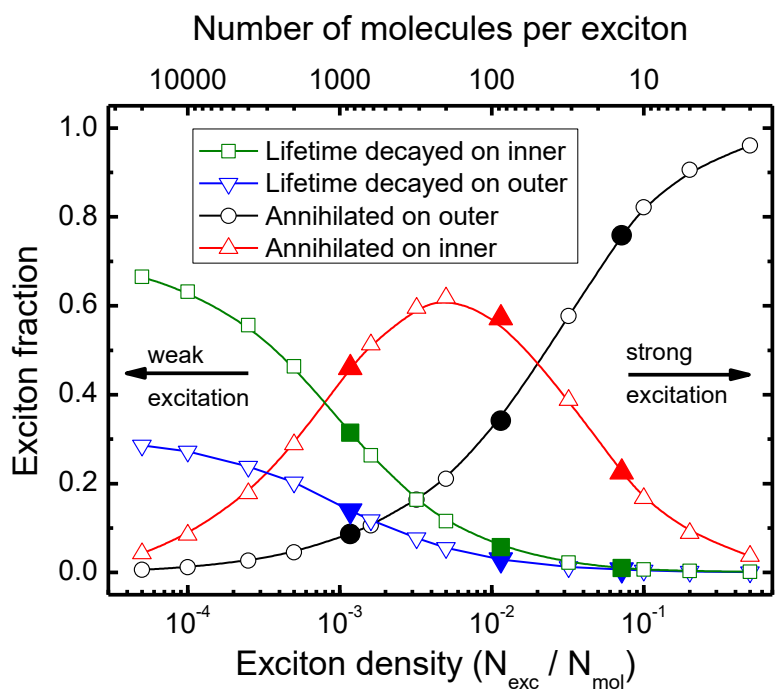

**Supplementary Figure 14.** Fraction of excitons that were originally planted on the outer tube that decay naturally on the inner tube (green squares), decay naturally on the outer tube (blue triangles), annihilate on the inner tube (red triangles) or annihilate on the outer tube (black circles) from the Monte-Carlo simulations as a function of the exciton density. At time zero both tubes are populated with the same

exciton density. The inverse exciton density, i.e., the number of molecules per exciton is plotted on the top axis. Solid symbols: exciton densities used in experiment on complete nanotubes; open symbols: additional data points for illustration of the trend. Solid lines are drawn to guide the eye of the reader.

In the following we analyze the distribution and the mean number of EEA events accumulated by individual excitons for which the corresponding histograms are shown in Supplementary Figure 15. Here, we evaluate the share of excitons that were planted on the outer tube, and died on either the inner (upper panels; red) or the outer tube (bottom panels; black) at the exciton densities used in experiment.

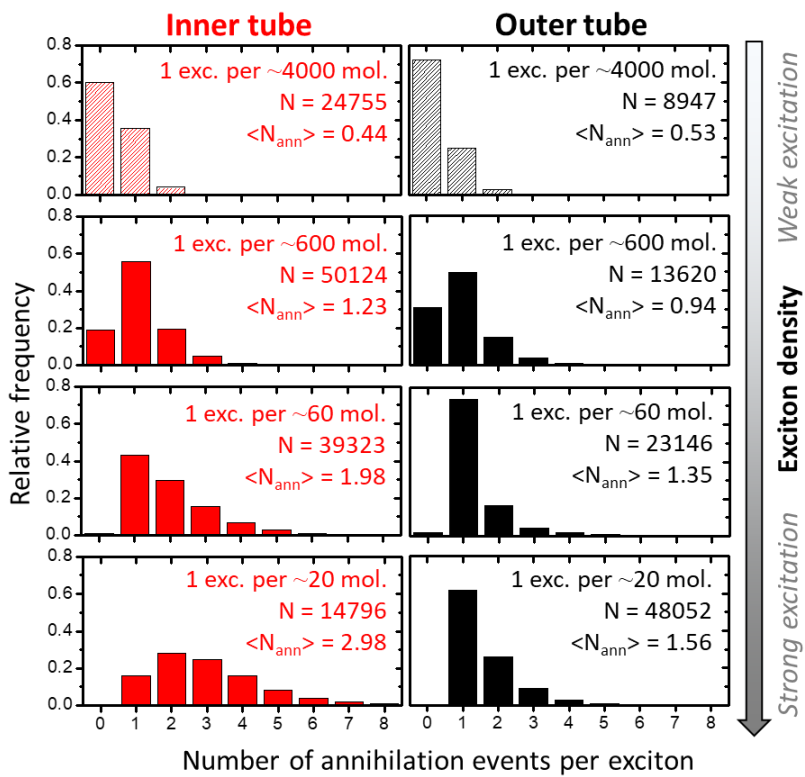

**Supplementary Figure 15.** Histograms of the accumulated number of annihilation events by individual excitons that were originally planted on the outer tube and died on the inner (red; left panels) or the outer tube (black; right panels) for different exciton densities. The upper panel does not refer to any exciton

density in experiment, but was added to illustrate the high probability of EEA even at exciton densities ~7 times lower than experimentally used. Inset: Total number of excitons for the respective histogram and mean number of annihilation events per exciton.

For higher exciton densities EEA gains importance at expense of excitons that decay naturally before engaging in any EEA event, i.e., zero annihilation events. Consequently, the distribution of the number of EEA participations gradually shifts to higher values, where the mean number of annihilation events dramatically increases (from 1.23 to 2.98) for the inner tube, while there is only a moderate increase (from 0.53 to 1.56) for the outer tubes. Note that the upper boundary for the mean value of 2 (as discussed in the previous section) is no longer applicable here, as the inner tube does not represent a closed system anymore, but can receive additional excitons from the outer tube. In the regime of high exciton densities, the first cascade of annihilation events occurs on the outer tube during which excitons participate in one or two EEA events ($< N_{\mathrm{ann}} >$ = 1.56). After the subsequent transfer of the surviving excitons to the inner tube, these excitons can engage in further EEA events, which leads to the large mean value of EEA participations on the inner layer, although EEA is more prominent on the outer tube in terms of total number of annihilated excitons.

## Supplementary Note 7: Observation of the Seventh-Order Signal

In the Monte-Carlo simulations, we implicitly included the occurrence of multi-exciton processes, where excitons could participate in multiple exciton–exciton annihilation events. These higher-order effects (in the language of nonlinear optics, seventh-order, *etc.*; see Supplementary Note 5.3) lead to additional changes of the observed dynamics of the absorptive and EEI signals. Experimentally the seventh-order signal could also be observed despite undersampling of the coherence time with the step size of 0.38 fs (see Methods section in the main text). The latter determined the Nyquist limit at about 44000 cm$^{-1}$, which lies below the expected position of the seventh-order signal at triple the fundamental frequency (~50000 cm$^{-1}$). However, due to back folding at the Nyquist limit the seventh-order signal appeared at (44000 - 6000) cm$^{-1}$ = 38000 cm$^{-1}$ along the excitation axis[8]. For high exciton densities this signal was indeed resolved in the EEI signal (Supplementary Figure 16), which corroborates the influence of multi-exciton processes for the experimentally observed signals as well as in our modelling.

The seventh-order signals are observed for both complete nanotubes as well as isolated inner tubes, but significantly stronger for the former. The structure of the seventh-order signals is identical to the lower-order signals with diagonal peaks for the outer and inner tube as well as a cross peak. The only difference is that the peak signs are again inverted compared to the EEI signal and, thus, identical to the absorptive signal, where the ground-state bleach (GSB) shows up negative and excited state absorption (ESA) positive. This again originates from the two additional interactions of the sample with the incident light fields in the perturbative expansion. A quantitative analysis of these signals, however, is hindered due to the low signal amplitudes as well as the multitude of involved processes and, therefore, is beyond the scope of this paper.

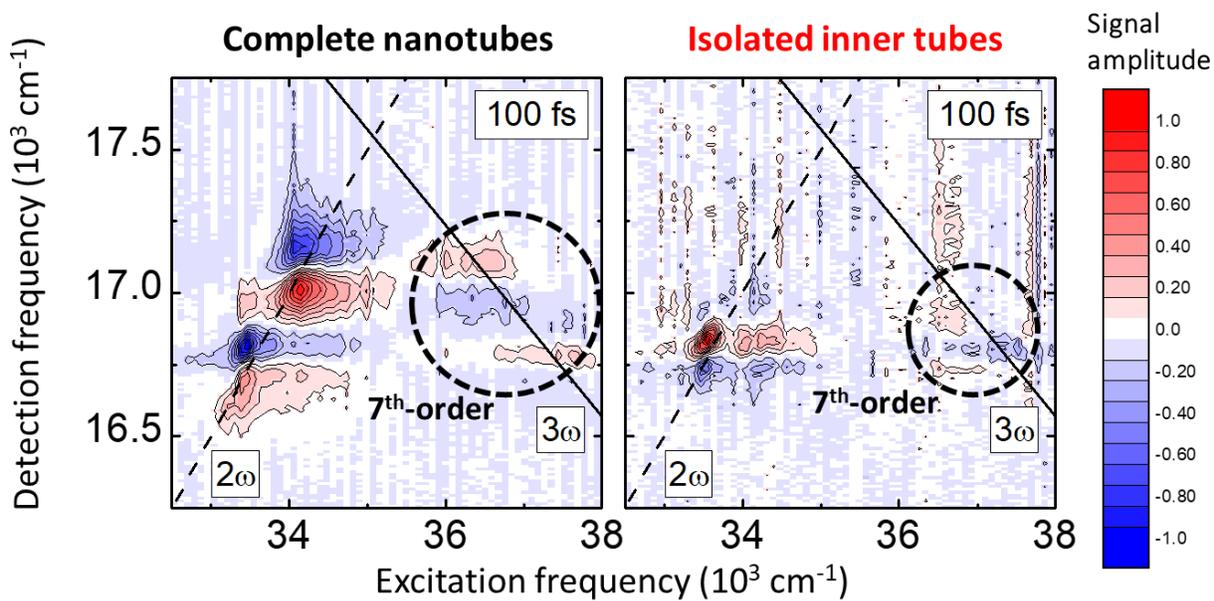

**Supplementary Figure 16**. EEI2D spectra of complete nanotubes (left) and isolated inner tubes (right) with the excitation axis extended to higher frequencies. The signal amplitude is depicted on a color scale. The 2D spectra were obtained at the highest exciton density in experiment for a waiting time of 100 fs. Dashed diagonal lines are drawn at $\omega_{excitation} = 2\omega_{detection}$. The anti-diagonal lines (solid) at $3\omega$ originate from back folding of the diagonal $\omega_{excitation} = 3\omega_{detection}$ line at the Nyquist limit (~44000 cm$^{-1}$).

# Supplementary Note 8: Diagonal Peak Absorptive and EEI Transients for Complete Nanotubes

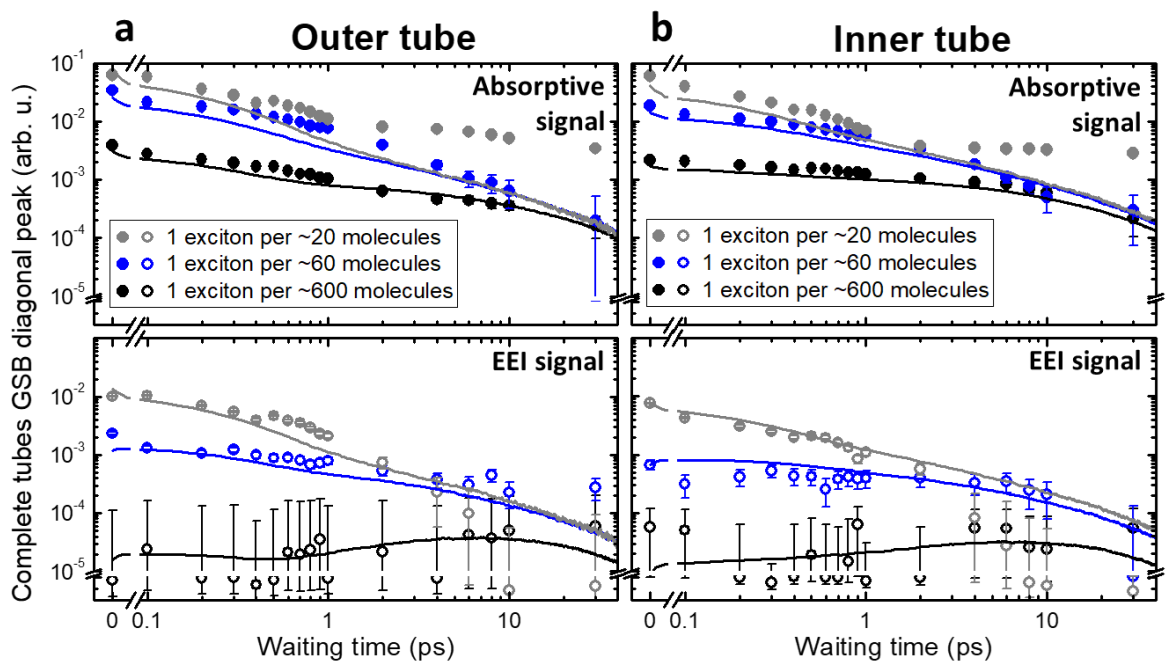

**Supplementary Figure 17.** Absorptive and EEI transients of both layers in case of complete nanotubes. Log-log plots of the absorptive (upper panels, solid circles) and EEI (lower panels, open circles) GSB/SE transients for (**a**) outer and (**b**) inner tube diagonal peaks at different exciton densities. The transients were obtained by integrating the signal in the rectangular regions of interest shown in Figure 2b in the main text. The panels are drawn with the same scaling to emphasize their direct comparability, as both are derived from the same signal. The error bars refer to the detection noise level in the experiment, i.e., the standard error of the background fluctuations in the respective spectral region during each measurement (Supplementary Note 2). The solid lines depict the results from Monte-Carlo simulations of the exciton dynamics on isolated inner tubes. The amplitude (vertical) scaling between experimental and simulated data is preserved, i.e., for each signal (absorptive and EEI) a single scaling factor was used for *all* simulated transients. The sign of the EEI responses was inverted for the ease of comparison. Deceleration

of the transient dynamics at T > 2 ps for the highest exciton density (1 exciton per ~20 molecules) is caused by transient heating of the nanotubes and a few surrounding water layers as a result of the energy released by exciton annihilation events (Supplementary Note 9).

## Supplementary Note 9: Thermal Heating Induced by Exciton–Exciton Annihilation

In the absorptive 2D spectra of complete nanotubes an interesting side effect of exciton–exciton annihilation was observed, which was reflected in transient heating of the nanotubes and a few surrounding solvent layers. The mechanism is the following: The energy of the annihilated exciton is transferred via a number of (vibrational) relaxation steps to low-frequency modes thereby creating a quasi-equilibrium Boltzmann distribution at elevated temperature. The whole relaxation process takes only a few ps which might be related to ultrafast cooling in liquid water[26,27]. The increased temperature leads to small but detectable modifications of the nanotubes' absorption spectrum, which causes a TA signal offset without detectable temporal variation on the time scale up to 100 ps especially evident at high exciton densities. The signal offset vanishes before the arrival of the next laser pulse, i.e. after 1 ms.

In order to investigate this effect in greater depth we have performed a series of transient absorption (TA) measurements with an extended scanning range of the delay time $T$. Here we indeed find a prominent signal offset for high exciton densities (a representative TA map is shown in Supplementary Figure 18a). In order to analyze the signal offset, we average the TA signal for delay times between 90 ps and 100 ps (Supplementary Figure 18a, side panel) and over 50 cm$^{-1}$ along the detection axis. Specifically, the intervals for averaging are 16660 cm$^{-1}$ to 16710 cm$^{-1}$ and 16950 cm$^{-1}$ to 17000 cm$^{-1}$ for the inner and outer tube, respectively. As a next step, the signal amplitudes are plotted as a function of the respective exciton density in experiment (Supplementary Figure 18b, circles). For comparison, the signal at zero delay time is extracted in the same spectral intervals and shown in the same graph (Supplementary Figure 18b, squares).

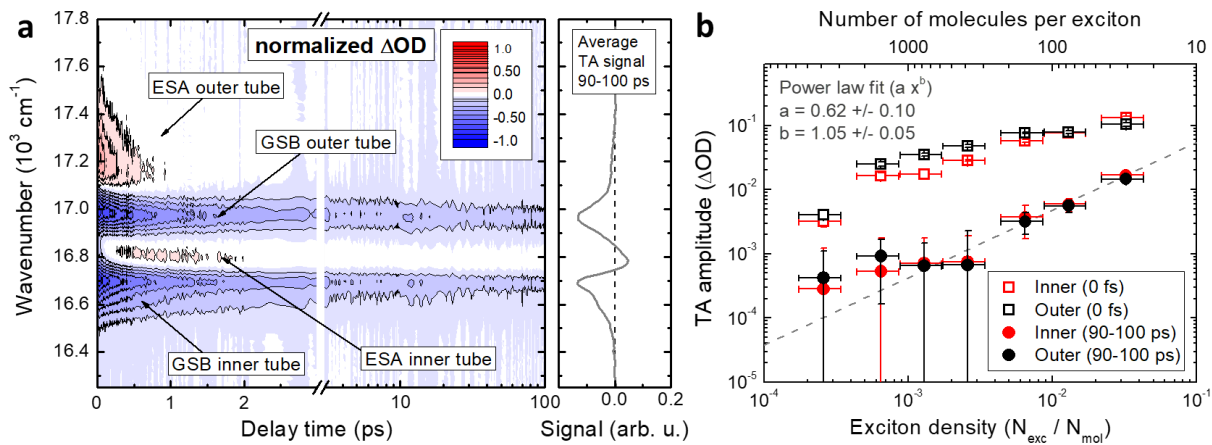

**Supplementary Figure 18.** (**a**) Transient absorption map of complete nanotubes for the highest exciton density (one exciton per ~30 molecules) normalized to the initial maximum absolute amplitude of the signal, i.e., change of optical density ($\Delta$OD). The latter is depicted on a color scale with increments of 0.1 and contour lines drawn as specified in the color bar. The vertical and horizontal axes are detection wavenumber and delay time between pump and probe pulse, respectively. The right panel depicts the average TA spectrum between 90 to 100 ps. (**b**) Log-log plot of the TA amplitude at delay times of 0 fs (open squares) and averaged between 90 ps and 100 ps (solid circles) for different exciton densities (average number of excitons per molecule) for the inner and outer tube in red and black, respectively. The top axis depicts the inverse exciton density, i.e., the number of molecules per one exciton, for simple

comparison with the main text. The error bars in the vertical direction refer to the standard error upon averaging multiple scans. The horizontal error bars depict the uncertainty of the exciton density.

The long-time offset of the TA signal (Supplementary Figure 18b, solid circles) scales linearly with the exciton density, while the truly non-linear, early-time signal has a saturation scaling (Supplementary Figure 18b, open squares). These are strong indications for heating of the sample, since the amount of energy dissipated into the system scales linearly with the excitation power. This linear scaling is evident from fitting of the offset amplitude with a power law ($f(x) = ax^b$; Supplementary Figure 18b, gray dashed line), which yields a near-unity exponent of $b = 1.05 \pm 0.05$. Furthermore, the amplitude of the offset exceeds the amplitude expected from a simple exponential decay of the TA signal, which indicates that the detected signal does not directly originate from the remaining exciton population at this delay time. According to the one-exciton lifetime of 33 ps for complete nanotubes, the signal is expected to decay to about 6 mOD (i.e., 5% of its initial amplitude ~130 mOD) after 100 ps, which is significantly lower than the measured value of ~17 mOD.

As a further test of this hypothesis we collected absorption spectra induced by a temperature jump, i.e., a change in temperature (black and orange curves in Supplementary Figure 19a). The spectra were measured by taking consecutive absorption spectra between which the sample was heated by $\Delta T = 2$ K and then allowed to slowly cool down to room temperature (RT) again. Throughout the measurement the sample temperature was monitored using a thermocouple submerged into the sample solution. The difference spectra were then computed as $\Delta \text{OD} = \text{OD}(\text{RT} + \Delta T) - \text{OD}(\text{RT})$ and depicted as black ($\Delta T = 2$ K) and orange ($\Delta T = 0.5$ K) curves in Supplementary Figure 19 in comparison with the TA "offset" spectrum at a delay time of 100 ps (blue).

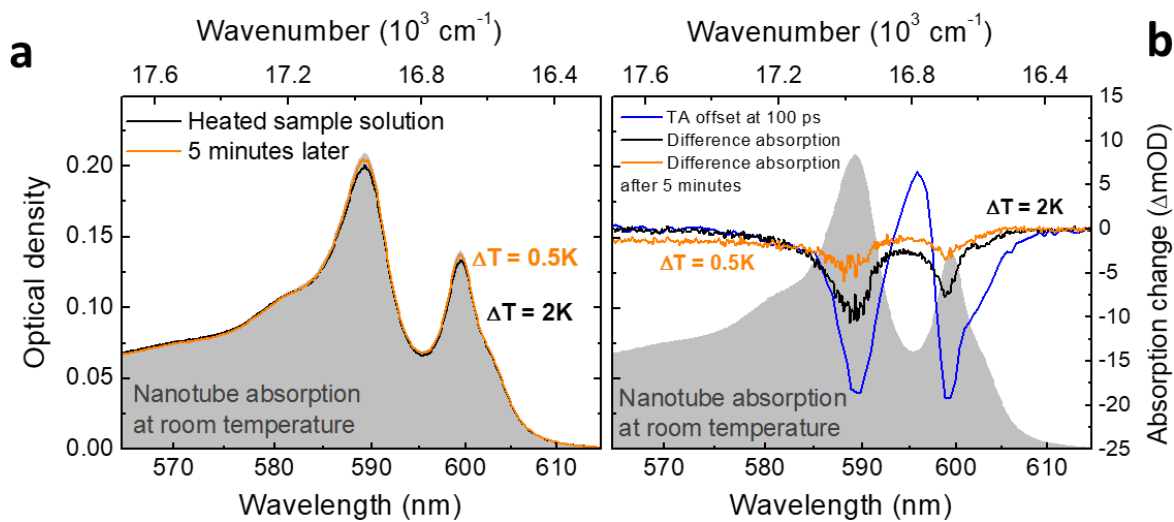

**Supplementary Figure 19.** (**a**) Absolute absorption spectra of complete nanotubes at room temperature (gray, shaded) and at increased temperatures ΔT = 2 K (black) and ΔT = 0.5 K (orange) after 5 min of waiting. (**b**) Pump–probe (TA) spectrum (blue) and difference spectra between C8S3 nanotube absorption spectra with a temperature difference of 2 K (black) and 0.5 K (orange).

Overall, the agreement of the TA offset and the difference spectra due to temperature regarding the spectral shape and peak positions is good with the exception of the magnitude being slightly underestimated so that a temperature change of 4 K would have likely been a better estimate. Nevertheless, these results strongly suggest that the offset observed in the absorptive 2D and TA experiments is likely to originate from transient heating of the sample.

A temperature change of 4 K corresponds to the heating due to a single laser shot (ΔE = 2.5 nJ) from which 15% are absorbed and subsequently converted into heat. To reach this estimate a number assumptions have to be made: (1) the heated volume is confined to the nanotube volume, (2) heat dissipation into the bulk solvent is negligible at the time scale of the experiment (~100 ps), and (3) the heat capacity of the nanotubes is identical to water (specific heat capacity $c_s$ =

4.1379 J g⁻¹ K⁻¹ at room temperature), as the heat capacity of the nanotubes is unknown. The change in temperature is then computed via

$$\Delta T = \frac{Q}{C} = \frac{A \, \Delta E}{c_s \, m_{H_2O}} = \frac{0.15 \times 2.5 \times 10^{-9} \, \text{J}}{4.1379 \frac{\text{J}}{\text{g K}} \times 2.2 \times 10^{-11} \, \text{g}} \approx 4 \, \text{K}$$

with the supplied heat $Q$, absorbed fraction of the laser pulse $A$, pulse energy $\Delta E$, and the heat capacity $C$. The corresponding mass of water was calculated from that fraction of the focal volume $V_{\text{foc}}$ that is actually occupied by nanotubes. Therefore, the latter were treated as simple hollow cylinders with an inner and outer radius of $R_{\text{inner}} = 3.551$ nm and $R_{\text{outer}} = 6.465$ nm, respectively. The focal volume is assumed cylindrical with a radius of 50 μm and thickness of 50 μm yielding for the water mass

$$m_{H_2O} = \left(\frac{c \, N_A \, V_{\text{foc}}}{\rho_{\text{total}}^{\text{mol}}}\right) \pi (R_{\text{outer}}^2 - R_{\text{inner}}^2) \rho_{H_2O} \approx 2.2 \times 10^{-11} \, \text{g}.$$

The first bracketed factor computes the total length of nanotubes in the focal volume via the molar concentration $c = 1.11 \times 10^{-4} \, M$ (Methods and Supplementary Note 1), Avogadro constant $N_A$, and the molar density $\rho_{\text{total}}^{\text{mol}} = 114$ nm⁻¹ (number of molecules per unit length of the nanotubes counting both layers; extracted from the theoretical model presented in Ref. [2]). Taken together with the cross section of the nanotube calculated from the inner and outer radii and the density of water ($\rho_{H_2O}$) yields the corresponding mass. Note that if the entire focal volume ($V_{\text{foc}}$) is assumed to heat up, the absolute change in temperature $\Delta T$ is in the sub-mK range.

# Supplementary Note 10: Photoluminescence (PL) Measurements for One-Exciton Lifetime

In order to accurately determine the one-exciton lifetime of complete nanotubes as well as flash-diluted inner tubes we measured the photoluminescence (PL) response of the sample at extremely low exciton densities of about one exciton per $10^4$ molecules for which the transients are shown in Supplementary Figure 20. In either case the sample was excited at 550 nm and the PL response was recorded with a streak camera (Hamamatsu, model C5680).

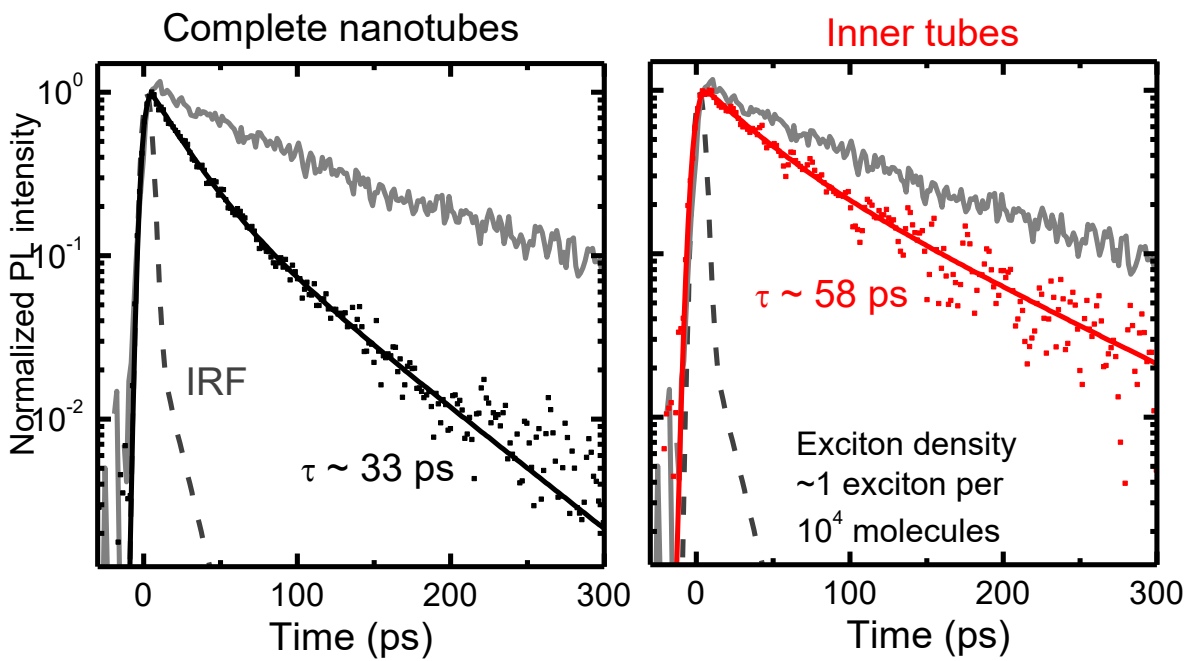

**Supplementary Figure 20.** Experimental PL transients for complete nanotubes (black dots) and isolated inner tubes (red dots) at low exciton densities of only one exciton per $10^4$ molecules. The excitation wavelength is 550 nm. The solid lines are fits according to a bi-exponential decay convoluted with the instrument response function (IRF, dashed). Gray line: PL decay of C8S3 monomers dissolved in methanol (lifetime ~100 ps).

Despite the low exciton density, we find that the transients exhibit a small degree of non-exponentiality. Therefore, we fit the transients with a convolution of two exponential decays and the instrument response function (IRF), which can be approximated by a Gaussian with standard deviation width of ~3 ps. From these fits we extract the weighted averaged lifetime of a single exciton for the isolated inner tubes and complete nanotubes as 58 ps and 33 ps, respectively. In either case the lifetime is shorter than for C8S3 monomers dissolved in methanol (~100 ps) due to the formation of a super-radiant state[5].

The lifetime of complete nanotubes is in good agreement with previously published values obtained from femtosecond transient grating photoluminescence measurements obtained for nanotubes suspended in a sugar matrix following 400 nm excitation[7]. It is worth noting that some studies reported PL lifetimes of ~64 ps at 100 K by freezing the nanotubes in their aqueous host solvent[23] and up to ~260 ps at room temperature for nanotubes suspended in a sugar matrix[28]. The cause for these differences is not exactly understood, although the choice of the host matrix as well as the experimental parameters such as excitation wavelengths (495 nm in Ref. [23]; 400 nm, and 520 nm in Ref. [28]) might play a role.

The reduced lifetime of complete nanotubes compared to the isolated inner tubes is consistent with earlier reported observations for matrix-suspended oxidized nanotubes[7]. We hypothesize that the presence of the outer tube may introduce additional non-radiative pathways through which inner-tube excitons decay.

# Supplementary Note 11: Exciton Diffusion Tensor using the Haken-Strobl-Reineker Model

## 11.1. Parametrization

For the molecular structure of C8S3 nanotubes, the same model (extended herringbone model with two molecules per unit cell) and parameters as reported in Eisele *et al.* are used[2]. To calculate the diffusion constant predicted by the Haken-Strobl-Reineker model of thermal fluctuations according to the expression given in the Methods section, it is necessary to know the exciton states and their corresponding energies. These are obtained for each one of the two walls separately by numerically diagonalizing the respective Hamiltonian:

$$H = \sum_{n=1}^{N} \varepsilon_n b_n^\dagger b_n + \sum_{n=1}^{N} \sum_{\substack{m=1 \\ n \neq m}}^{N} J_{nm} b_n^\dagger b_m$$

Here, $\varepsilon_n$ corresponds to the excitation energy of molecule n, which is taken from a Gaussian distribution with mean $\varepsilon_o = 18868$ cm$^{-1}$ and standard deviation $\sigma = 250$ cm$^{-1}$ in order to account for static disorder. These energetic parameters are the same as previously reported in Ref. [2]. Furthermore, $b_n^\dagger (b_n)$ denote the Pauli operators[2] for the creation (annihilation) of an excitation on molecule *n*. The intermolecular couplings $J_{nm}$ are calculated using extended dipole-dipole interactions[2] and are assumed to be non-fluctuating quantities. The number of molecules is set to $N_{\text{inner}} = 7992$ for the inner wall and $N_{\text{outer}} = 14260$ for the outer wall, which corresponds to tubes with almost equal length of approximately $L \approx 197$ nm in accordance with Ref. [2]. From these values a molecular surface density of 1.81 molecules nm$^{-2}$ was extracted as $\rho = N/2\pi RL$, where $N$ is the total number of molecules in each tube and $R$ is the radius of the tube. Furthermore, the inter-tube interactions are neglected in the simulation. Under this premise the only two remaining parameters to calculate the diffusion constant tensor are the dephasing rate $\Gamma$ that characterizes the thermal white noise fluctuations of the Haken-Strobl-Reineker model and the temperature $T$. Here, we use $\Gamma = 83.75$ cm$^{-1}$, which represents the average of the Lorentzian lineshape widths (half width half maximum) used in Ref. [2] and $T = 295$ K, i.e., room temperature. We note that as seen in Supplementary Figure 21 the diffusion constants are only weakly dependent on $\Gamma$ in the relevant parameter regime.

### 11.2. Diffusion Tensor Elements

As a result of the calculation, we obtain the diffusion tensor elements shown in Supplementary Figure 21 as a function of the dephasing rate $\Gamma$.

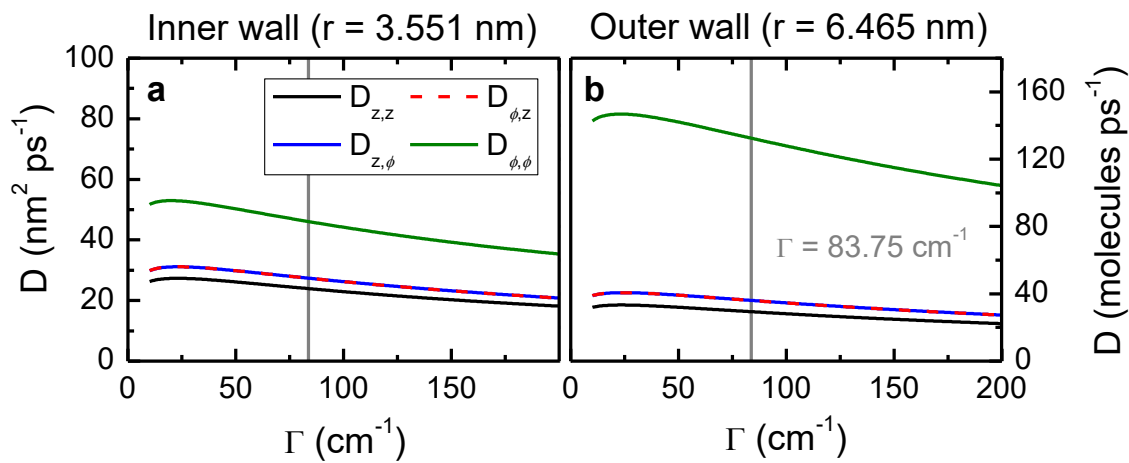

**Supplementary Figure 21.** Diffusion tensor elements as a function of the dephasing rate $\Gamma$ for (**a**) the inner and (**b**) outer wall of double-walled nanotubes in units of nm$^2$ ps$^{-1}$ (left axis) and molecules ps$^{-1}$

(right axis). The temperature in calculations was set to 295 K and the static Gaussian energy disorder strength σ = 250 cm$^{-1}$. The vertical line (gray) marks the dephasing rate $\Gamma$ relevant to the nanotubes.

We use the axial component ($D_{z,z}$) of the diffusion constant tensor for comparison to the results from our experiments and Monte-Carlo simulations. We justify this choice by the high aspect ratio of the nanotubes (i.e., ~20 in the MC simulations) for which $D_{z,z}$ is expected to be the dominant component for exciton–exciton annihilation in particular for longer diffusion times, whereas exciton diffusion around the tube ($D_{\phi,\phi}$) is less important. This assumption holds for any exciton density, as excitons on the perimeter would rapidly annihilate after which the later dynamics are again governed by the axial diffusion constant. The different elements of the diffusion constant tensor at a given dephasing rate of $\Gamma = 83.75$ cm$^{-1}$ (HWHM) are summarized in Supplementary Table 5.

**Supplementary Table 5.** Individual elements of the diffusion constant tensor for a dephasing rate of $\Gamma = 83.75$ cm$^{-1}$ (HWHM) for the inner and outer tube.

|  | **Inner tube** | **Outer tube** |
|---|---|---|
| $D_{z,z}$ | 23.9 nm$^2$ps$^{-1}$ | 16.3 nm$^2$ps$^{-1}$ |
| $D_{\phi,z} = D_{z,\phi}$ | 27.4 nm$^2$ps$^{-1}$ | 20.0 nm$^2$ps$^{-1}$ |
| $D_{\phi,\phi}$ | 46.0 nm$^2$ps$^{-1}$ | 73.6 nm$^2$ps$^{-1}$ |

## Supplementary Note 12: Monte-Carlo Simulations for Purely Diffusive Exciton Dynamics

Here we test a scenario in which the exciton–exciton interaction radius is nullified, i.e., two excitons only annihilate in case they occupy the same site after a hopping event, which is compensated by an increased hopping rate to obtain a diffusion constant of 100 nm$^2$ ps$^{-1}$ in accordance with previously published results[30]. All other conditions as outlined in the main text, stayed unaltered.

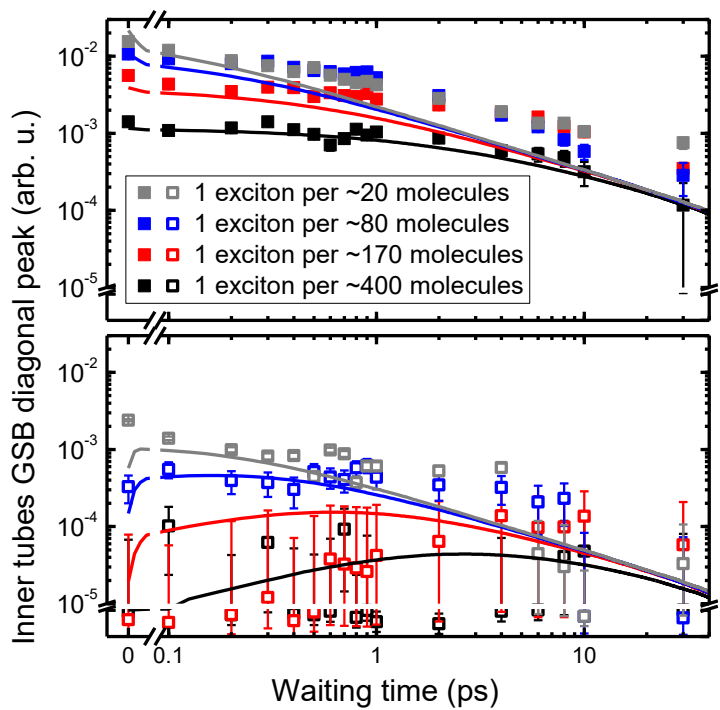

**Supplementary Figure 22.** Experimental absorptive and EEI transients for isolated inner tubes as a function of waiting time (symbols). The experimental data are the same as in Figure 3 in the main paper. The solid lines depict transients from MC simulations.

In the regime of high exciton densities (Supplementary Figure 22, gray), the dynamics at early waiting times are captured reasonably well, as a faster exciton diffusion can compensate the lack of an extended annihilation radius and *vice versa*. However, towards longer waiting times the increased diffusion constant leads to unsatisfactory fit of the data, as the calculated dynamics are generally too fast. In particular, this trend becomes apparent at intermediate exciton densities, where the simulations predicts the maximum EEI signal to occur at ~600 fs (Supplementary Figure 22, red) and ~200 fs (Supplementary Figure 22, blue), although the experimental data reach the maximum amplitude at ~6 ps and ~1 ps, respectively. Therefore, we conclude that exciton diffusion alone cannot account for the experimental observations, but an extended radius for exciton–exciton interactions is required to describe the data adequately.

## Supplementary Note 13: EEI2D Setup Schematic

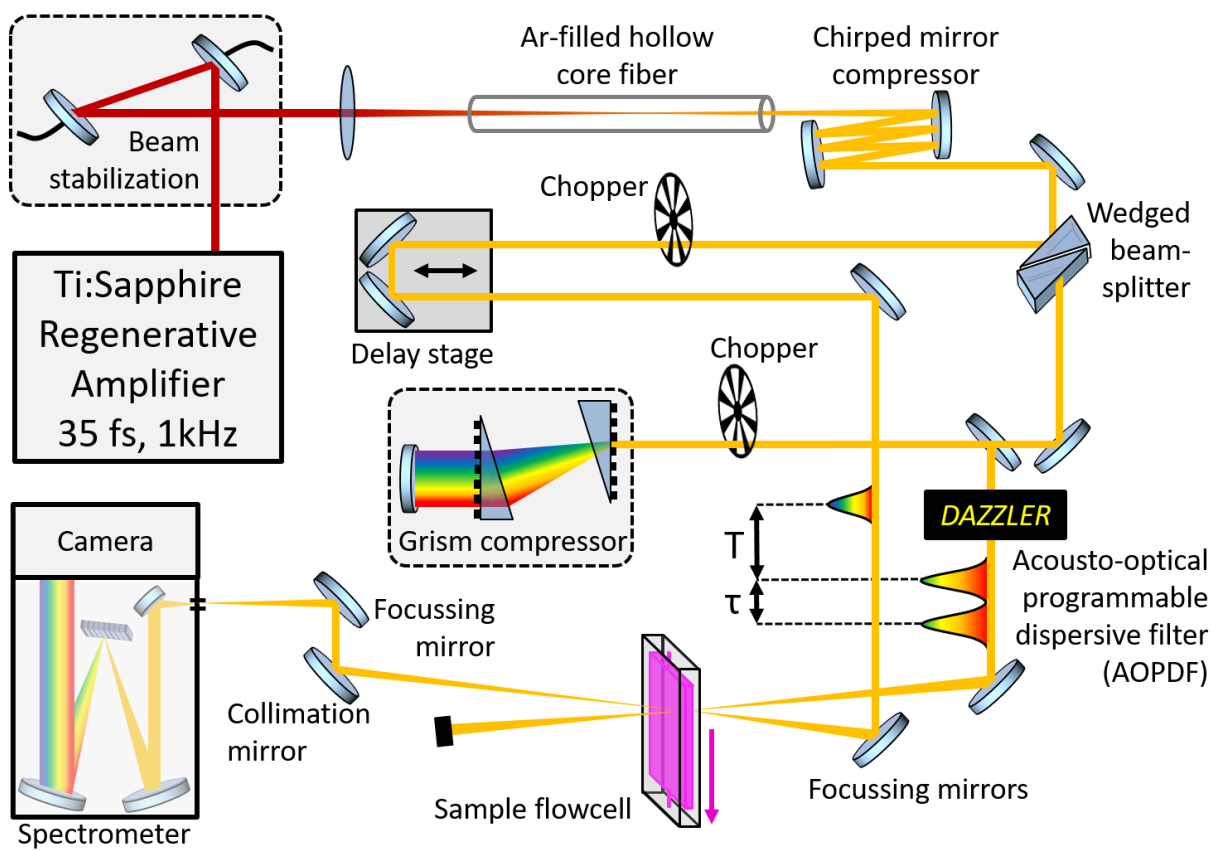

**Supplementary Figure 23.** Schematic of the experimental apparatus for absorptive 2D and EEI2D spectroscopy. The flow direction of the sample is indicated by the arrow. The pump and probe pulses are polarized along the flow.

## Supplementary Note 14: EEI2D Spectra on Laser Dye Sulforhodamine 101

As a control experiment, absorptive 2D and EEI2D spectra of the laser dye sulforhodamine 101 (SR101, Radiant Dyes) diluted in water at concentration of $10^{-4}$ mol L$^{-1}$ were recorded (Supplementary Figure 24) at which the average distance between individual molecules amounts to ~26 nm. SR101 was chosen as its absorption peak is located in the same spectral range as the nanotube absorption spectrum investigated here. The response from diluted SR101 molecules is expected to be annihilation-free, as the individual molecules are spaced far apart and, thus, energetically uncoupled. Therefore, any photo-excitation remains localized on a single molecule, which prevents exciton–exciton annihilation. The optical density was set at OD ≈ 0.08 at 586 nm, which is similar to the OD of the nanotubes sample; the excitation energy was set at 40 nJ per pulse. This resulted in excitation of approximately 10% of the SR101 molecules in the focal volume, or, in terms of the main text, one excitation per 10 molecules. This exceeds the highest exciton density used for 2D spectroscopy of the nanotubes by a factor of ~2.

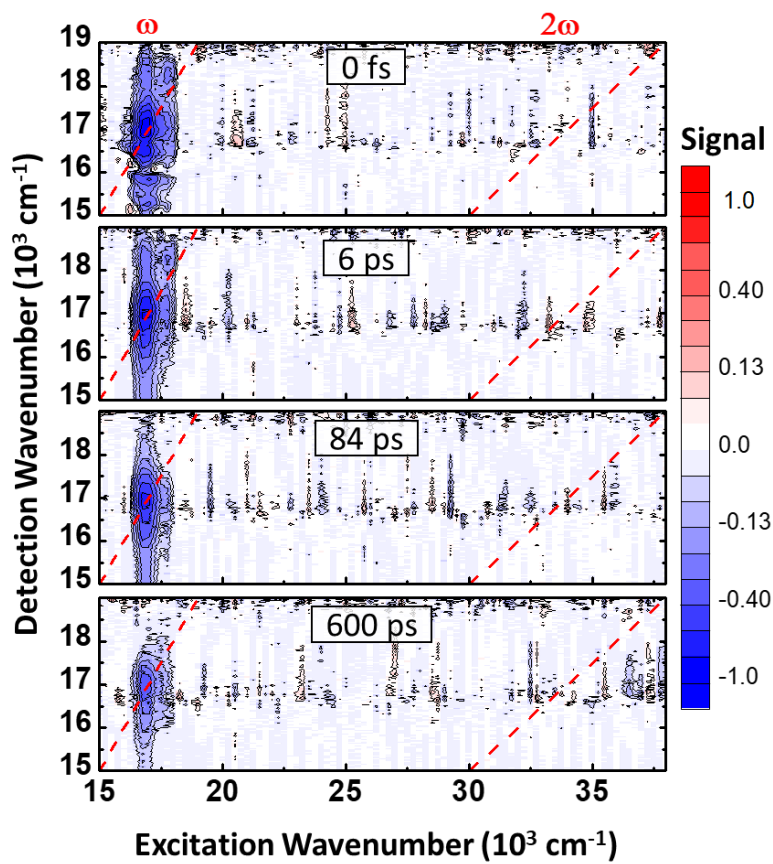

**Supplementary Figure 24.** Absorptive 2D and EEI2D spectra of sulforhodamine 101 dissolved in $H_2O$ for different waiting times for an excitation power of 40 nJ. The signal amplitude was normalized to the maximum absolute amplitude (at 0 fs waiting time). Dashed lines (red) are drawn at $\omega_{excitation} = \omega_{detection}$ and $\omega_{excitation} = 2\omega_{detection}$ to mark the positions of the absorptive and EEI signals, respectively.

The 2D absorptive response is governed by a broad (negative) ground-state bleach GSB and stimulated emission (SE) signal around the fundamental frequency $\omega$, while there is no EEI signal detectable around the double frequency $2\omega$. Furthermore, Supplementary Figure 24 depicts the full range of excitation frequencies from 15000 cm$^{-1}$ to 36000 cm$^{-1}$ in order to prove that the signal at intermediate frequencies is free of any artifacts or spurious signals. These results confirm that the spectral range, where EEI signals are expected, is free from artifacts from the experimental apparatus. Importantly, it also supports our assignment of the EEI signal arising from exciton–exciton annihilation, because an even higher excitation density for dissolved sulforhodamine 101 molecules does not result in any observable EEI signal, which in turn justifies our theoretical approach.

**Supplementary Note 15: Inter-Wall Excitation Transfer Rate**

In order to determine the transfer rate from the outer to the inner tube for complete nanotubes, absorptive 2D spectroscopy was used with fine sampling of the waiting time in steps of 5 fs. To minimize the influence of exciton–exciton annihilation, the exciton density was set to only one exciton per ~500 molecules, which caused the increased noise level of the transient. Supplementary Figure 25 depicts the ground-state bleach (GSB) cross peak amplitude as a function of waiting time.

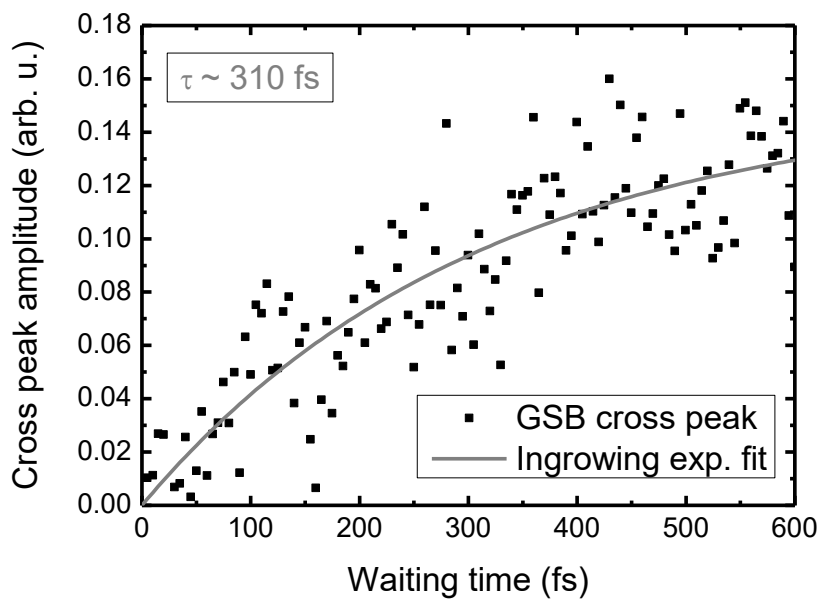

**Supplementary Figure 25.** GSB cross peak amplitude of complete nanotubes as a function of waiting time (black squares). The gray line is an exponential fit of the transient.

Fitting the GSB cross peak transient with an in-growing exponential function $F(t) = A\,(1 - \exp(-t/\tau))$ yields a time constant of ~310 fs, which is in excellent agreement with the values reported in literature[7,21,22].

## Supplementary References

1. Dostál, J.; Fennel, F.; Koch, F.; Herbst, S.; Würthner, F.; Brixner, T. Direct observation of exciton–exciton interactions. *Nat. Commun.* **9,** 2466 (2018).

2. Eisele, D. M.; Cone, C. W.; Bloemsma, E. A.; Vlaming, S. M.; van der Kwaak, C. G. F.; Silbey, R. J.; Bawendi, M. G.; Knoester, J.; Rabe, J. P.; Vanden Bout, D. A. Utilizing redox-chemistry to elucidate the nature of exciton transitions in supramolecular dye nanotubes. *Nat. Chem.* **4,** 655–662 (2012).

3. Kriete, B.; Bondarenko, A. S.; Jumde, V. R.; Franken, L. E.; Minnaard, A. J.; Jansen, T. L. C.; Knoester, J.; Pshenichnikov, M. S. Steering Self-Assembly of Amphiphilic Molecular Nanostructures via Halogen Exchange. *J. Phys. Chem. Lett.* **8,** 2895–2901 (2017).

4. Clark, K. A.; Cone, C. W.; Vanden Bout, D. A. Quantifying the Polarization of Exciton Transitions in Double-Walled Nanotubular J-Aggregates. *J. Phys. Chem. C* **117,** 26473–26481 (2013).

5. Spano, F. C.; Mukamel, S. Superradiance in molecular aggregates. *J. Chem. Phys.* **91,** 683–700 (1989).

6. Igor, S.; Tobias, B.; Mino, Y.; Fleming, G. R. Heterogeneous Exciton Dynamics Revealed by Two-Dimensional Optical Spectroscopy. *J. Phys. Chem. B* **110,** 20032–20037 (2006).

7. Pandya, R.; Chen, R.; Cheminal, A.; Thomas, T. H.; Thampi, A.; Tanoh, A.; Richter, J. M.; Shivanna, R.; Deschler, F.; Schnedermann, C.; *et al.* Observation of Vibronic Coupling Mediated Energy Transfer in Light-Harvesting Nanotubes Stabilized in a Solid-State

Matrix. *J. Phys. Chem. Lett.* **9,** 5604–5611 (2018).

8.  Hamm, P.; Zanni, M. *Concepts and Methods of 2D Infrared Spectroscopy*. (Cambridge University Press, 2011).

9.  Mukamel, S. *Principles of Nonlinear Optical Spectroscopy*. (Oxford University Press, Oxford, 1995).

10. Süß, J.; Wehner, J.; Dostál, J.; Brixner, T.; Engel, V. Mapping of exciton–exciton annihilation in a molecular dimer via fifth-order femtosecond two-dimensional spectroscopy. *J. Chem. Phys.* **150,** 104304 (2019).

11. Malý, P.; Mančal, T. Signatures of Exciton Delocalization and Exciton–Exciton Annihilation in Fluorescence-Detected Two-Dimensional Coherent Spectroscopy. *J. Phys. Chem. Lett.* **9,** 5654–5659 (2018).

12. von Berlepsch, H.; Kirstein, S.; Hania, R.; Pugzlys, A.; Böttcher, C.; Pugžlys, A.; Böttcher, C.; Pugzlys, A.; Böttcher, C. Modification of the nanoscale structure of the J-aggregate of a sulfonate-substituted amphiphilic carbocyanine dye through incorporation of surface-active additives. *J. Phys. Chem. B* **111,** 1701–1711 (2007).

13. Qiao, Y.; Polzer, F.; Kirmse, H.; Kirstein, S.; Rabe, J. P. Nanohybrids from nanotubular J-aggregates and transparent silica nanoshells. *Chem. Commun.* **51,** 11980–11982 (2015).

14. Didraga, C.; Pugžlys, A.; Hania, P. R.; von Berlepsch, H.; Duppen, K.; Knoester, J. Structure, spectroscopy, and microscopic model of tubular carbocyanine dye aggregates. *J. Phys. Chem. B* **108,** 14976–14985 (2004).

15. Sperling, J.; Nemeth, A.; Hauer, J.; Abramavicius, D.; Mukamel, S.; Kauffmann, H. F.; Milota, F. Excitons and disorder in molecular nanotubes: a 2D electronic spectroscopy study and first comparison to a microscopic model. *J. Phys. Chem. A* **114,** 8179–8189 (2010).

16. Megow, J.; Röhr, M. I. S. S.; Schmidt am Busch, M.; Renger, T.; Mitrić, R.; Kirstein, S.; Rabe, J. P.; May, V.; am Busch, M.; Renger, T.; *et al.* Site-dependence of van der Waals interaction explains exciton spectra of double-walled tubular J-aggregates. *Phys. Chem. Chem. Phys.* **17,** 6741–7 (2015).

17. Czikkely, V.; Försterling, H. D.; Kuhn, H. Light absorption and structure of aggregates of dye molecules. *Chem. Phys. Lett.* **6,** 11–14 (1970).

18. Engel, E.; Leo, K.; Hoffmann, M. Ultrafast relaxation and exciton–exciton annihilation in PTCDA thin films at high excitation densities. *Chem. Phys.* **325,** 170–177 (2006).

19. Fennel, F.; Lochbrunner, S. Exciton-exciton annihilation in a disordered molecular system by direct and multistep Förster transfer. *Phys. Rev. B* **92,** 140301 (2015).

20. Förster, T. Zwischenmolekulare energiewanderung und fluoreszenz. *Ann. Phys.* **437,** 55–75 (1948).

21. Augulis, R.; Pugzlys, A.; van Loosdrecht, P. H. M.; Pugžlys, A.; van Loosdrecht, P. H. M.; Pugzlys, A.; van Loosdrecht, P. H. M. Exciton dynamics in molecular aggregates. *Phys. status solidi* **3,** 3400 (2006).

22. Yuen-Zhou, J.; Arias, D. H.; Eisele, D. M.; Steiner, C. P.; Krich, J. J.; Bawendi, M. G.; Nelson, K. A.; Aspuru-Guzik, A. Coherent exciton dynamics in supramolecular light-harvesting nanotubes revealed by ultrafast quantum process tomography. *ACS Nano* **8,** 5527–5534 (2014).

23. Pugžlys, A.; Augulis, R.; Van Loosdrecht, P. H. M.; Didraga, C.; Malyshev, V. A.; Knoester, J. Temperature-dependent relaxation of excitons in tubular molecular aggregates: fluorescence decay and Stokes shift. *J. Phys. Chem. B* **110,** 20268–20276 (2006).

24. Clark, K. A.; Krueger, E. L.; Vanden Bout, D. A. Temperature-Dependent Exciton Properties of Two Cylindrical J-Aggregates. *J. Phys. Chem. C* **118,** 24325–24334 (2014).

25. Mikhnenko, O. V.; Blom, P. W. M.; Nguyen, T.-Q. Exciton diffusion in organic semiconductors. *Energy Environ. Sci.* **8,** 1867–1888 (2015).

26. Yeremenko, S.; Pshenichnikov, M. S.; Wiersma, D. A. Interference effects in IR photon echo spectroscopy of liquid water. *Phys. Rev. A* **73,** 021804 (2006).

27. Lindner, J.; Vöhringer, P.; Pshenichnikov, M. S.; Cringus, D.; Wiersma, D. A.; Mostovoy, M. Vibrational relaxation of pure liquid water. *Chem. Phys. Lett.* **421,** 329–333 (2006).

28. Caram, J. R.; Doria, S.; Eisele, D. M.; Freyria, F. S.; Sinclair, T. S.; Rebentrost, P.; Lloyd, S.; Bawendi, M. G. Room-Temperature Micron-Scale Exciton Migration in a Stabilized Emissive Molecular Aggregate. *Nano Lett.* **16,** 6808–6815 (2016).

29. Davydov, A. S. *Theory of Molecular Excitons*. (Plenum, 1971).

30. Clark, K. A.; Krueger, E. L.; Vanden Bout, D. A. Direct measurement of energy migration in supramolecular carbocyanine dye nanotubes. *J. Phys. Chem. Lett.* **5,** 2274–2282 (2014).